\documentclass[12pt]{iopart}
\usepackage{hyperref,bm}
\begin{document}
\title[Weyssenhoff fluid dynamics in general relativity using a 1+3 covariant approach]{Weyssenhoff fluid dynamics in general relativity using a 1+3 covariant approach}
\author{S D Brechet, M P Hobson, A N Lasenby}
\address{Astrophysics Group, Cavendish Laboratory, J.~J.~Thomson Avenue, Cambridge, CB3 0HE, UK}
\eads{\mailto{sdb41@mrao.cam.ac.uk}, \mailto{mph@mrao.cam.ac.uk}, \mailto{a.n.lasenby@mrao.cam.ac.uk}}
\begin{abstract}
The Weyssenhoff fluid is a perfect fluid with spin where the spin of the matter fields is the source of torsion in an Einstein-Cartan framework. Obukhov and Korotky showed that this fluid can be described as an effective fluid with spin in general relativity. A dynamical analysis of such a fluid is performed in a gauge invariant manner using the $1+3$ covariant approach. This yields the propagation and constraint equations for the set of dynamical variables. A verification of these equations is performed for the special case of irrotational flow with zero peculiar acceleration by evolving the constraints.
\end{abstract}
\submitto{\CQG}
\maketitle

\section{Introduction}
The Einstein-Cartan theory of gravity (EC) extends Einstein's theory of general relativity (GR) in a natural way by including the spin properties of matter and their influence on the geometrical structure of space-time. By removing the symmetry requirement on the two lower indices of the connection, Cartan $\cite{Cartan:1922}$ showed that the dynamics is no longer entirely determined by the metric; the antisymmetric part of the connection called torsion became an independent dynamical variable. Besides the energy-momentum of the matter content sourcing curvature, its spin was later postulated to be the source of torsion $\cite{Hehl:1973}$. The EC theory locally satisfies the Poincar\'e symmetry $\cite{Hehl:1976}$ accounting for translational degrees of freedom associated with curvature and rotational degrees of freedom linked to torsion.

Weyssenhoff and Raabe initiated a careful study of the behaviour of perfect fluids with spin $\cite{Weyssenhoff:1947}$ . In order to build cosmological models based on the EC theory, Obukhov and Korotky extended their work $\cite{Obukhov:1987}$. They showed, in particular, that by assuming the Frenkel condition$\footnote{Note that the Frenkel condition arises naturally when performing a rigorous variation of the action. It simply means that the spin pseudovector is spacelike in the fluid rest frame.}$ the model reduces to the description of an effective fluid in GR where the effective stress-energy momentum tensor contains some additional spin squared terms.

The first studies of perturbations of a perfect fluid within GR were carried out by Lifshitz $\cite{Lifshitz:1946}$ in a fixed gauge and reformulated in terms of gauge-invariant variables by Bardeen $\cite{Bardeen:1980}$. The dynamics of such a fluid have also been investigated in a more physical and transparent gauge-invariant manner by Hawking $\cite{Hawking:1966}$ and extended by Ellis $\cite{Ellis:1989}$. We shall follow the latter approach here and use the $1+3$ formalism.

As Puetzfeld points out $\cite{Puetzfeld:2004}$, there are an increasing number of theoretical reasons for studying cosmological models based on a non-Riemannian geometry, as some key features of the current concordance model such as dark matter, dark energy and in particular inflation still need to be explained. The Weyssenhoff fluid, for example, seems a promising candidate to describe cosmological inflation in a geometrical manner without using scalar fields, which have not yet been observed. This promising behaviour may arise from the spin density squared terms contained within the effective stress energy momentum tensor derived by Obukhov and Korotky $\cite{Obukhov:1987}$, since these spin contributions dominate the dynamics at early times. Although the Weyssenhoff fluid is expected to leave the late time dynamics unchanged, making it an unsuitable candidate to describe dark energy, it may still therefore significantly affect the early time evolution of the fluid.

In this publication, we restricted our study to the formal derivation of the dynamical relations for a Weyssenhoff fluid. A detailed study of the large scales dynamics of such a fluid in an attempt to get a spin based inflation will be pursued in further work. To remain as general as possible we chose not to perform a first- or second-order perturbation analysis for a particular class of models. This can easily be done according to the symmetries of the models, and some specific examples will be pursued in a later publication. The derivation of the Weyssenhoff fluid dynamics is a prelude to the perturbation analysis, which is especially relevant to study the structure formation seeded during the inflationnary era. The dynamics of such a fluid in a $1+3$ covariant approach has been studied previously in a cosmological context by Palle $\cite{Palle:1998}$. However, the use of effective GR relations in conjunction with EC identities is rather opaque in this work, and also certain length scales are excluded from the analysis making a new study, which considers all length scales, appropriate.

In the standard GR theory, the $1+3$ covariant approach leads to six propagation equations and six constraint equations. These give respectively the time and spatial covariant derivatives of the set of dynamical variables, which are the energy density $\rho$, the expansion rate $\Theta$, the shear density $\sigma$, the vorticity density $\omega$, the `electric' part of the Weyl tensor $E$ and the `magnetic' part of the Weyl tensor $H$. The Weyssenhoff fluid is described by an effective GR theory, where the additional degrees of freedom due to torsion are entirely determined by the spin density $S$. Therefore, in addition to the spin density modifying the dynamical equations for the six standard variables, we also expect to find additional dynamical relations.

In the next section, we briefly outline the EC theory, then give a concise description of a Weyssenhoff fluid in $\Sref{Section 3}$. $\Sref{Section 4}$ is devoted to the Weyssenhoff fluid dynamical analysis using the $1+3$ formalism outlined in {\it Appendix} A. The consistency of the particular case with zero vorticity and peculiar acceleration ($\omega=a=0$) is established by evolving the constraints in $\Sref{Section 5}$. The last section draws a comparison with Palle's results. In this paper, we use the $(+,-,-,-)$ signature. To express our results in the opposite signature used by Ellis $\cite{Ellis:1998}$, the correspondence between physical variables can be found in $\cite{Challinor:2000}$ and in {\it Appendix} B.

\section{Einstein-Cartan theory}
In the EC theory, the effect of the spin density tensor is locally to induce torsion in the structure of space-time. In holonomic coordinates, the torsion tensor ${Q^{\lambda}}_{\mu\nu}$ is defined as the antisymmetric part of the affine connection ${{\tilde{\Gamma}}^{\lambda}}_{\ \mu\nu}$,
\begin{equation}
{Q^{\lambda}}_{\mu\nu}={\tilde{\Gamma}^{\lambda}}_{\ [\mu\nu]}=
{\textstyle\frac{1}{2}}\left({\tilde{\Gamma}^{\lambda}}_{\
\mu\nu}-{\tilde{\Gamma}^{\lambda}}_{\ \nu\mu}\right)\
,\label{torsion def}
\end{equation}
which vanishes in GR since the connection is assumed to be symmetric in its two lower indices. Note that the tilde denotes an EC geometrical object to differentiate it from an effective GR object. In the following, Greek indices refer to a holonomic coordinate basis, while Latin indices refer to an arbitrary non-holonomic orthonormal basis.

In order to find a proper description of a Weyssenhoff fluid, we first have to determine the EC field equations. The gauge group associated with the EC theory is the Poincar\'e group $\cite{Hehl:1976}$. This is easy to understand as the asymmetry of the connection requires an affine generalisation of the Lorentz group which is precisely the Poincar\'e group. In the Poincar\'e gauge theory of gravity, the gravitational field is described by the tetrad field ${e_{\mu}}^{a}$ and the local spin connection ${\tilde{\omega}^{ab}}_{\ \ \mu}$. The spin connection is antisymmetric in its Latin indices, ${\tilde{\omega}^{ab}}_{\ \ \mu}=-{\tilde{\omega}^{ba}}_{\ \ \mu}$ if $\tilde{\nabla}_{\lambda}g_{\mu\nu}=0$ which we assume throughout, and the inverse of the tetrad is given by ${e^{\mu}}_{a}$, such that ${e^{\mu}}_{a}{e_{\mu}}^{b}=\delta^{\,b}_{\,a}$ and ${e^{\mu}}_{a}{e_{\nu}}^{a}=\delta^{\,\mu}_{\,\nu}$. The geometrical structure of $U_4$ $\--$ i.e. the metric $g_{\mu\nu}$ and the EC connection ${\tilde{\Gamma}^{\lambda}}_{\ \mu\nu}$ $\--$ is completely determined by the tetrad (translational field) and the spin connection (rotational field) according to,
\begin{eqnarray}
g_{\mu\nu}&={e_{\mu}}^{a}{e_{\nu}}^{b}\eta_{ab}\ ,\label{tetradmetric}\\
{\tilde{\Gamma}^{\lambda}}_{\ \mu\nu}&=
{e^{\lambda}}_{a}{\tilde{\omega}^{a}}_{\
b\nu}{e_{\mu}}^{b}+{e^{\lambda}}_{a}\partial_{\nu}{e_{\mu}}^{a}\
.\label{conspin}
\end{eqnarray}
Using the gauge relations $\eref{tetradmetric}$ and $\eref{conspin}$, the torsion tensor $\eref{torsion def}$ can be rewritten in terms of the translational and rotational fields,
\begin{equation}
{Q^{a}}_{\mu\nu}={e_{\lambda}}^{a}{{\tilde{\Gamma}^{\lambda}}}_{\
[\mu\nu]}=-{\textstyle\frac{1}{2}}\left(
\partial_{\mu}{e_{\nu}}^{a}-\partial_{\nu}{e_{\mu}}^{a}
+{\tilde{\omega}^{a}}_{\ b\mu}{e_{\nu}}^{b}-{\tilde{\omega}^{a}}_{\
b\nu}{e_{\mu}}^{b}\right)\ .\label{torsiongauge}
\end{equation}
The metric and the connection are assumed to be compatible, which means that the nonmetricity vanishes and implies that the EC connection $\tilde{\Gamma}^{\lambda}_{\ \mu\nu}$ can be decomposed in terms of the Levita-Civita (torsion free) connection $\Gamma^{\lambda}_{\ \mu\nu}$ and the contortion tensor $K^{\lambda}_{\ \mu\nu}$ as,
\begin{equation}
{\tilde{\Gamma}^{\lambda}}_{\ \mu\nu}=\Gamma^{\lambda}_{\
\mu\nu}-K^{\lambda}_{\ \mu\nu}\ ,\label{torsionconnection}
\end{equation}
where,
\begin{eqnarray}
{\Gamma^{\lambda}}_{\mu\nu}&={\textstyle\frac{1}{2}}g^{\lambda\sigma}
(\partial_{\mu}g_{\sigma\nu}+\partial_{\nu}g_{\mu\sigma}-\partial_{\sigma}g_{\mu\nu})\
,\label{Levi-Civita con}\\
K^{\lambda}_{\
\mu\nu}&=-{Q^{\lambda}}_{\mu\nu}-{Q_{\mu\nu}}^{\lambda}-{Q_{\nu\mu}}^{\lambda}\
.\label{Contortion}
\end{eqnarray}
The curvature is described by the Riemann-Cartan tensor and its contractions, i.e. the Ricci-Cartan tensor and the Ricci-Cartan scalar,
\begin{eqnarray}
&\tilde{{R^a}}_{b\mu\nu}=\partial_{\mu}{\tilde{\omega}^{a}}_{\ b\nu}-\partial_{\nu}{\tilde{\omega}^{a}}_{\ b\mu} +{\tilde{\omega}^{c}}_{\ b\nu}{\tilde{\omega}^{a}}_{\ c\mu}-{\tilde{\omega}^{c}}_{\ b\mu}{\tilde{\omega}^{a}}_{\ c\nu}\
,\label{Riemanntorsion}\\
&\tilde{R}_{\mu\nu}={\tilde{R}^{\sigma}}_{\ \mu\sigma\nu}
={e^{\sigma}}_{a}{e_{\mu}}^{b}{\tilde{R}^{a}}_{\
b\sigma\nu}\ ,\label{Riccitorsion}\\
&\tilde{\mathcal{R}}={\tilde{R}^{\sigma\nu}}_{\ \ \sigma\nu}
={e^{\sigma}}_{a}{e^{\nu}}_{c}\eta^{cb}{\tilde{R}^{a}}_{\
b\sigma\nu}\ .\label{RicciScaltorsion}
\end{eqnarray}
The field equations of the EC theory are derived from the action $S$ defined on a space-time manifold $\mathcal{M}$ as,
\begin{equation}
S=\int_{\mathcal{M}}d^4x\left[\frac{e}{2\kappa}\left(\tilde{\mathcal{R}}-2\Lambda\right)+\mathcal{L}_m\right]
,\label{action torsion}
\end{equation}
where $\kappa=8\pi G/c^4$, $e=\mathrm{det}({e_{\mu}}^{a})$, $\Lambda$ is the cosmological constant and $\mathcal{L}_m=\mathcal{L}_m({e_{\mu}}^{a},\tilde{\omega}^{ab}_{\ \ \mu},\phi_m)$ is the Lagrangian density of the matter fields $\phi_m$ . Varying the action $\eref{action torsion}$ independently for ${e_{\mu}}^{a}$ and $\tilde{\omega}^{ab}_{\ \ \mu}$, the field equations are respectively found to be,
\begin{eqnarray}
\tilde{R}^{\mu}_{\
a}-{\textstyle\frac{1}{2}}{e^{\mu}}_{a}\tilde{\mathcal{R}}+{e^{\mu}}_{a}\Lambda=\kappa\tilde{T}^{\mu}_{\
a}\
,\label{trans fieldeq}\\
Q^{\mu}_{\ ab}+2{e^{\mu}}_{[a\vphantom]}Q_{\vphantom[b]}=\kappa
S^{\mu}_{\ ab}\ ,\label{rot fieldeq}
\end{eqnarray}
where $Q_a={Q^{\mu}}_{a\mu}$ is the torsion trace, and the material sources of the gravitational field are respectively the energy-momentum and the spin density tensors defined as,
\begin{eqnarray}
{\tilde{T}^{\mu}}_{\ \ a}\equiv\frac{1}{e}\frac{\delta
\mathcal{L}_m}{\delta {e_{\mu}}^{a}}\
,\label{energy-momentum}\\
S^{\mu}_{\ ab}\equiv\frac{1}{e}\frac{\delta
\mathcal{L}_m}{\delta\tilde{\omega}^{ab}_{\ \ \mu}}\ .\label{spin}
\end{eqnarray}
These source terms are the functional tensors of the EC classical field theory obtained by variation of the action $S$. They should not be confused with the corresponding canonical tensors derived from Noether's theorem since these two kinds of tensors may differ in an EC framework. The translational field equation $\eref{trans fieldeq}$ can be recast in terms of purely holonomic coordinates and decomposed into symmetric and anti-symmetric parts,
\begin{eqnarray}
\tilde{R}_{(\mu\nu)}-{\textstyle\frac{1}{2}}g_{\mu\nu}\tilde{\mathcal{R}}
+g_{\mu\nu}\Lambda=\kappa\tilde{T}_{(\mu\nu)}\
,\label{trans sym}\\
\tilde{R}_{[\mu\nu]}=\kappa\tilde{T}_{[\mu\nu]}\ .\label{trans
antisym}
\end{eqnarray}

\section{Weyssenhoff fluid description}
\label{Section 3}

The Weyssenhoff fluid is a continuous macroscopic medium which is characterized on microscopic scales by the spin of the matter fields. The spin density of matter is described by an antisymmetric tensor,
\begin{equation}
S_{\mu\nu}=-S_{\nu\mu}\ ,\label{Spin density tens}
\end{equation}
and has been postulated by Obukhov and Korotky $\cite{Obukhov:1987}$ to be related to the source of torsion according to,
\begin{equation}
{S^{\lambda}}_{\mu\nu}=u^{\lambda}S_{\mu\nu}\ ,\label{Spin density
tens II}
\end{equation}
where $u^{\lambda}$ is the $4$-velocity of the fluid element. The Frenkel condition requires the intrinsic spin of a matter field to be spacelike in the rest frame of the fluid,
\begin{equation}
S_{\mu\nu}u^{\nu}=0\ .\label{Frenkel}
\end{equation}
This condition arises naturally from a rigorous variation of the matter Lagrangian $\mathcal{L}_m$ as shown in $\cite{Obukhov:1987}$.

The Frenkel condition implies that the torsion trace vanishes, and hence the rotational field equations $\eref{rot fieldeq}$ reduce to an algebraic coupling between spin and torsion according to,
\begin{equation}
Q^{\lambda}_{\ \mu\nu}=\kappa u^{\lambda}S_{\mu\nu}\
.\label{algebraic coupling}
\end{equation}

Thus, the torsion contributions to the EC field equations are entirely described in terms of the spin density. It is useful to introduce a spin-density scalar $S$ defined as,
\begin{equation}
S^2={\textstyle\frac{1}{2}}S_{\mu\nu}S^{\mu\nu}\geq 0\ .\label{Spin density
scalar}
\end{equation}

Using the Frenkel condition, Obukhov and Korotky showed $\cite{Obukhov:1987}$ that the symmetric part of the EC field equations for a perfect fluid with spin $\eref{trans sym}$ can be recast in terms of effective GR Einstein field equations with additional spin terms, whereas the antisymmetric part $\eref{trans antisym}$ simply becomes a GR spin field equation.

The former are found to be,
\begin{equation}
R_{\mu\nu}-{\textstyle\frac{1}{2}}g_{\mu\nu}\mathcal{R}=\kappa T^{s}_{\mu\nu}\ ,
\label{Einstein ef eq}
\end{equation}
where the effective stress energy momentum tensor of the fluid is given by,
\begin{equation}
T^{s}_{\mu\nu}=(\rho_s+p_s)u_{\mu}u_{\nu}-p_sg_{\mu\nu}
-2\left(g^{\rho\lambda}+u^{\rho}u^{\lambda}\right)\nabla_{\rho}\left[u_{(\mu\vphantom)}S_{\vphantom(\nu)\lambda}\right]\ , \label{Ef stress en tensor}
\end{equation}
with effective energy density and pressure of the form,
\begin{eqnarray}
\eqalign{\rho_s=\rho-\kappa S^2+\kappa^{-1}\Lambda\ , \\
p_s=p-\kappa S^2-\kappa^{-1}\Lambda\ ,}\label{Ef energy pressure}
\end{eqnarray}
satisfying the physical equation of state,
\begin{equation}
p=w\rho\ ,\label{equation of state}
\end{equation}
where $w$ is the equation of state parameter.

The spin field equation is given by,
\begin{equation}
\nabla_{\lambda}\left(u^{\lambda}S_{\mu\nu}\right)
=2u^{\rho}u_{[\mu\vphantom]}\nabla_{|\lambda}\left(u^{\lambda}S_{\vphantom[\rho|\nu]}\right)\
. \label{Ef spin field equations}
\end{equation}

\section{Weyssenhoff fluid dynamics using a 1+3 covariant approach}
\label{Section 4}

We will now use the $1+3$ covariant approach, outlined for convenience in {\it Appendix} A, to describe accurately the dynamics of a Weyssenhoff fluid in GR on all scales and in a non-perturbative way. Once the dynamical evolution is entirely determined, a perturbation analysis can be performed for any given class of models according to their symmetries. In a cosmological context, we would require the cosmological fluid to be highly symmetric on large scales but allow for generic inhomogeneities on small scales. This is necessary to provide an accurate enough description of the observable universe accounting for its homogeneity and isotropy on large scales as well as for all the complicated structures it contains on small scales.

In GR, the Weyssenhoff fluid dynamics is actually a generalisation of the dynamics of a perfect fluid, where the effective energy density $\rho_s$ and pressure $p_s$ contain a spin density squared $S^2$ correction term, and the stress energy momentum tensor $T^{s}_{\mu\nu}$ incorporates an additional spin divergence term. The new contribution to the effective dynamics comes from the spin field equation $\eref{Ef spin field equations}$. Thus, the dynamics of a perfect fluid is recovered for a vanishing spin density.

The dynamical model of a perfect fluid with spin is fully determined by its matter content $\--$ including the spin properties of the particles $\--$ and its curvature. The matter content of the Weyssenhoff fluid is described by the effective stress-energy momentum tensor $\eref{Ef stress en tensor}$. Using the $1+3$ formalism, it can be recast as,
\begin{eqnarray}
\eqalign{
T_{\mu\nu}^{s}=&\left(\rho_s+4\omega^{\lambda}S_{\lambda}\right)u_{\mu}u_{\nu}-p_sh_{\mu\nu}\\
&-2u_{(\mu\vphantom)}D^{\lambda}S_{\vphantom(\nu)\lambda}
+4u_{(\mu\vphantom)}a^{\lambda}S_{\vphantom(\nu)\lambda}
-2{\sigma_{(\mu\vphantom)}}^{\lambda}S_{\vphantom(\nu)\lambda}
+2{\omega_{(\mu\vphantom)}}^{\lambda}S_{\vphantom(\nu)\lambda}\
.}\label{stress energy mom 1+3}
\end{eqnarray}
The physical interpretation of the Weyssenhoff fluid now becomes more transparent. The terms containing the effective energy density $\rho_s$ and pressure $p_s$ represent the behaviour of an effective perfect fluid, where $\rho_s$ and
$p_s$ account for the spin contributions. The other terms describe how the peculiar acceleration of the fluid $a_{\mu}$ and the fluid anisotropies $\--$ described by the rate-of-shear $\sigma_{\mu\nu}$ and the vorticity $\omega_{\mu\nu}$ respectively $\--$ couple to the spin density $S_{\mu\nu}$ and contribute to the effective energy density of the fluid.

All the information related to the curvature is encoded in the Riemann tensor which can be decomposed as $\cite{Hawking:1966}$,
\begin{equation}
R^{\rho\mu}_{\ \ \ \nu\lambda}=C^{\rho\mu}_{\ \ \ \nu\lambda} -
\delta^{\rho}_{\ [\lambda\vphantom]}R^{\mu}_{\ \vphantom[\nu]} -
\delta^{\mu}_{\ [\nu\vphantom]}R^{\rho}_{\ \vphantom[\lambda]} -
{\textstyle\frac{1}{3}}\mathcal{R}\delta^{\rho}_{\ [\nu\vphantom]}\delta^{\
\mu}_{\ \ \vphantom[\lambda]}\ ,\label{Riemann tens}
\end{equation}
where $C^{\rho\mu}_{\ \ \ \nu\lambda}$ is the Weyl tensor constructed to be the trace-free part of the Riemann tensor.

By analogy to classical electrodynamics, the Weyl tensor can be split relative to $u^{\mu}$ into an `electric' and a `magnetic' part $\cite{Hawking:1966}$ according to,
\begin{eqnarray}
E_{\mu\nu} = C_{\mu\rho\nu\sigma}u^{\rho}u^{\sigma}\ ,\label{Elec}\\
H_{\mu\nu} =\vphantom{}^{\ast}C_{\mu\rho\nu\sigma}u^{\rho}u^{\sigma} =
{\textstyle\frac{1}{2}}\eta_{\mu\sigma\lambda}C^{\sigma\lambda}_{\ \ \
\nu\rho}u^{\rho}\ ,\label{Magn}
\end{eqnarray}
where $\vphantom{}^{\ast}C_{\mu\nu\rho\sigma}$ is the dual of the Weyl tensor. These parts represent the `free gravitational field', enabling gravitational action at a distance and describing tidal forces and gravitational waves.

The Ricci tensor $R_{\mu\nu}$ is simply obtained by substituting the expression $\eref{stress energy mom 1+3}$ for the effective stress energy momentum tensor $T_{\mu\nu}^{s}$ into the Einstein field equations $\eref{Einstein ef eq}$,
\begin{eqnarray}
\eqalign{ R_{\mu\nu}=\kappa\Big\{&{\textstyle\frac{1}{2}}\left(\rho_s+3p_s+8\omega^{\lambda}S_{\lambda}\right)u_{\mu}u_{\nu}
-{\textstyle\frac{1}{2}}\left(\rho_s-p_s\right)h_{\mu\nu}\\
&-2u_{(\mu\vphantom)}D^{\lambda}S_{\vphantom(\nu)\lambda}
+4u_{(\mu\vphantom)}a^{\lambda}S_{\vphantom(\nu)\lambda}
-2{\sigma_{(\mu\vphantom)}}^{\lambda}S_{\vphantom(\nu)\lambda}
+2{\omega_{(\mu\vphantom)}}^{\lambda}S_{\vphantom(\nu)\lambda}\big\}\ .}\label{Ef Ricci tensor}
\end{eqnarray}

The Riemann tensor $R^{\rho\mu}_{\ \ \ \nu\lambda}$ can be fully split in a $1+3$ manner according to $\eref{Riemann tens}$ by using the expression $\eref{Ef Ricci tensor}$ for the Ricci tensor $R_{\mu\nu}$ and the decomposition of the Weyl tensor $C^{\rho\mu}_{\ \ \ \nu\lambda}$ into its electric $E_{\mu\nu}$ and magnetic $H_{\mu\nu}$ parts. For convenience, the tensor is split into three parts: the spinning perfect fluid part (P), the electric part of the Weyl tensor (E) and the magnetic part of the Weyl tensor (H). The decomposition yields,
\begin{equation}
R^{\rho\mu}_{\ \ \nu\lambda}\ =\ R^{\rho\mu}_{P\ \nu\lambda} +
R^{\rho\mu}_{E\ \nu\lambda} + R^{\rho\mu}_{H\ \nu\lambda}\
,\label{Riemann decomp}
\end{equation}
where
\begin{eqnarray*}
\fl \qquad R^{\rho\mu}_{P\ \nu\lambda}\ = &{\textstyle\frac{2}{3}}\kappa\left(\rho_s +
3p_s+12\omega^{\lambda}S_{\lambda}\right)h^{[\rho\vphantom]}_{\ \
[\nu\vphantom]}u^{\vphantom[\mu]}u_{\vphantom[\lambda]}
- {\textstyle\frac{2}{3}}\kappa\rho_sh^{[\rho\vphantom]}_{\ \ [\nu\vphantom]}h^{\vphantom[\mu]}_{\ \ \vphantom[\lambda]}\\
&-2\kappa\left({h^{[\rho\vphantom]}}_{[\nu\vphantom]}-u^{[\rho\vphantom]}u_{[\nu\vphantom]}\right)
[-u^{\vphantom[\mu]}D^{\sigma}S_{\vphantom[\lambda]\sigma}-u_{\vphantom[\lambda]}D_{\sigma}S^{\vphantom[\mu]\sigma}
+2u^{\vphantom[\mu]}a^{\sigma}S_{\vphantom[\lambda]\sigma}+2u_{\vphantom[\lambda]}a_{\sigma}S^{\vphantom[\mu]\sigma}\vphantom]\\
&\ \vphantom[\phantom{\ -2\kappa\left({h^{[\rho\vphantom]}}_{[\nu\vphantom]}-u^{[\rho\vphantom]}u_{[\nu\vphantom]}\right)}
-\sigma^{\vphantom[\mu]\sigma}S_{\vphantom[\lambda]\sigma}-\sigma_{\vphantom[\lambda]\sigma}S^{\vphantom[\mu]\sigma}
+\omega^{\vphantom[\mu]\sigma}S_{\vphantom[\lambda]\sigma}+\omega_{\vphantom[\lambda]\sigma}S^{\vphantom[\mu]\sigma}]
\ ,\\
\fl \qquad R^{\rho\mu}_{E\ \nu\lambda}\ = &C^{\rho\mu}_{E\ \nu\lambda}\ =\
4u^{[\rho\vphantom]}u_{[\nu\vphantom]}E^{\vphantom[\mu]}_{\ \
\vphantom[\lambda]} - 4h^{[\rho\vphantom]}_{\ \
[\nu\vphantom]}E^{\vphantom[\mu]}_{\ \ \vphantom[\lambda]}\
,\\
\fl \qquad R^{\rho\mu}_{H\ \nu\lambda}\ = &C^{\rho\mu}_{H\ \nu\lambda}\ =\
2\eta^{\rho\mu\sigma}u_{[\nu\vphantom]}H_{\vphantom[\lambda]\sigma}
+
2\eta_{\nu\lambda\sigma}u^{[\rho\vphantom]}H^{\vphantom[\mu]\sigma}\
.
\end{eqnarray*}

Note that for a vanishing spin density (i.e.\ in absence of torsion), we recover Ellis and van Elst's results $\cite{Ellis:1998}$ after reexpressing the physical variables in terms of the opposite signature $(-,+,+,+)$. This is also the case for every propagation and constraint equation describing the dynamics of the Weyssenhoff fluid because these expressions are projections of effective GR identities which are based on the Riemann tensor and its contractions.

In general, there are four sets of dynamical equations for a perfect fluid with spin. These sets are derived respectively from the Ricci identities, the Bianchi identities, once- and twice-contracted, and the spin field equation. We now discuss each set in turn.

\subsection{Ricci identities}

The first set of dynamical equations arises from the Ricci identities for the vector field $u^{\mu}$ defining the worldline of every matter field, i.e.,
\begin{equation}
2\nabla_{[\mu\vphantom]}\nabla_{\vphantom[\nu]}u_{\rho}=R_{[\mu\nu]\rho
}^{\ \ \ \ \ \lambda}u_{\lambda}\
.\label{Ricci identities}
\end{equation}
To extract the physical information stored in the Ricci identities, the latter have to be projected along the worldlines $u^{\mu}$ and on the orthogonal spatial hypersurfaces $h^{\mu}_{\ \nu}$. The non-vanishing projections yield the propagation equations and the constraint equations respectively,
\begin{eqnarray}
u^{\alpha}h^{\beta}_{\ \mu}h^{\gamma}_{\
\nu}\left(2\nabla_{[\alpha\vphantom]}\nabla_{\vphantom[\beta]}u_{\gamma}
-R_{[\alpha\beta]\gamma
}^{\ \ \ \ \ \sigma}u_{\sigma}\right)= \,0\ ,\label{Ricci prop}\\
\eta^{\rho\lambda}_{\ \ \ \nu}h^{\alpha}_{\ \rho}h^{\beta}_{\
\lambda}h^{\gamma}_{\
\mu}\left(2\nabla_{[\alpha\vphantom]}\nabla_{\vphantom[\beta]}u_{\gamma}
-R_{[\alpha\beta]\gamma
}^{\ \ \ \ \ \sigma}u_{\sigma}\right)= \,0\ ,\label{Ricci const}
\end{eqnarray}
where the latter have been expressed in terms of rank-$2$ tensors by duality ($\eta^{\rho\lambda}_{\ \ \ \nu}$) without loss of information.

The Ricci identities can be further split by separating the propagation and constraint equations into their trace part (T), symmetric trace-free part (STF) and antisymmetric trace-free part (ATF). The sets of equations are explicitly determined by the kinematics of the $1+3$ covariant formalism $\eref{kinematics}$ and by substituting the Riemann tensor decomposition $\eref{Riemann decomp}$ into the projections yielding the propagation $\eref{Ricci prop}$ and constraint $\eref{Ricci const}$ equations respectively before splitting them into parts.

\vspace{0.5cm} The propagation equations are found to be as follows.

\begin{itemize}
\item The Raychaudhuri equation (T),
\begin{equation}
\dot{\Theta}=-{\textstyle\frac{1}{3}}\Theta^2+D_{\lambda}a^{\lambda}+2\left(\omega^2-\sigma^2-a^2\right)
-{\textstyle\frac{\kappa}{2}}\left(\rho_s + 3p_s +
8\omega^{\lambda}S_{\lambda}\right)\ ,\label{Raychaudhuri eq}
\end{equation}
which is the basic dynamical equation of a perfect fluid with spin in this system. The last term on the RHS describes how the interaction between the spin density and the vorticity density affects the large scale dynamics. The physical meaning of this term is clear: the energy required to align the spin with the vorticity will act like a brake on the expansion, leading to the presence of this damping term in the Raychaudhuri equation.

\item The vorticity propagation equation (ATF),
\begin{equation}
\dot{\omega}_{\langle\mu\rangle}=-{\textstyle\frac{2}{3}}\Theta\,\omega_{\mu}+{\textstyle\frac{1}{2}}\left(\mathrm{curl}\
a\right)_{\mu}+\sigma_{\mu}^{\ \lambda}\omega_{\lambda}\
,\label{Vorticity prop eq}
\end{equation}
which shows how vorticity conservation follows for a perfect fluid. Note that there is no spin contribution, which means that torsion does not explicitly affect the vorticity evolution, although the effect of spin on the other dynamical variables must be taken into account.

\item The shear propagation equation (STF),
\begin{eqnarray}
\eqalign{
\dot{\sigma}_{\langle\mu\nu\rangle}=&-{\textstyle\frac{2}{3}}\Theta\,\sigma_{\mu\nu}+D_{\langle\mu}a_{\nu\rangle}-a_{\langle\mu}a_{\nu\rangle}
-\sigma_{\langle\mu}^{\ \ \lambda}\sigma_{\nu\rangle\lambda}+
\omega_{\langle\mu}\omega_{\nu\rangle}-E_{\mu\nu}\\
&+\kappa\left(\sigma_{\langle\mu}^{\ \
\lambda}S_{\nu\rangle\lambda}-\omega_{\langle\mu}S_{\nu\rangle}\right)
,}\label{Shear prop eq}
\end{eqnarray}
which shows how the tidal gravitational field $E_{\mu\nu}$ and the spin density $S_{\mu\nu}$ induce shear. The coupling between the spin density and the shear density contributes to the fluid anisotropies by increasing the rate of shear whereas the coupling between the spin density and the vorticity density has the opposite effect.
\end{itemize}

\vspace{0.5cm} The constraint equations are given by the following
relations.

\begin{itemize}
\item The vorticity divergence constraint (T),
\begin{equation}
D_{\lambda}\omega^{\lambda}=-a_{\lambda}\omega^{\lambda}\
.\label{Vorticity divergence id}
\end{equation}
This constraint simply expresses the fact that, in presence of a peculiar acceleration induced by a non-gravitational force due to the fluid dynamics, the spatial variation of vorticity is proportional to the vorticity.

\item The shear and spin divergence constraint (ATF),
\begin{equation}
D_{\lambda}\left({\sigma_{\mu}}^{\lambda}+{\omega_{\mu}}^{\lambda}+\kappa S_{\mu}^{\
\lambda}\right)-{\textstyle\frac{2}{3}}D_{\mu}\Theta=2a_{\lambda}\left({\omega_{\mu}}^{\lambda}+\kappa
{S_{\mu}}^{\lambda}\right)\ .\label{Shear div eq}
\end{equation}
Using the vorticity constraint $\eref{Vorticity divergence id}$, the shear and spin density constraint $\eref{Shear div eq}$ can be recast as,
\begin{equation}
D_{\lambda}\left({\sigma_{\mu}}^{\lambda}+\kappa S_{\mu}^{\
\lambda}\right)-{\textstyle\frac{2}{3}}D_{\mu}\Theta=a_{\lambda}\left(3{\omega_{\mu}}^{\lambda}+2\kappa
{S_{\mu}}^{\lambda}\right)\ .\label{Shear div recast}
\end{equation}
This expression relates the spatial variation of physical quantities, such as the spin density, the rate of shear and the expansion rate on the LHS, to the coupling between the acceleration due to the fluid dynamics and the fluid anisotropies on the RHS.

\item The magnetic constraint (STF),
\begin{equation}
H_{\mu\nu}=-D_{\langle\mu}\omega_{\nu\rangle}+2a_{\langle\mu}\omega_{\nu\rangle}
+\left(\mathrm{curl}\ \sigma\right)_{\mu\nu} .\label{Magnentic eq}
\end{equation}
Using the vorticity constraint $\eref{Vorticity divergence id}$, the magnetic constraint $\eref{Magnentic eq}$ reduces to,
\begin{equation}
H_{\mu\nu}=3a_{\langle\mu}\omega_{\nu\rangle}
+\left(\mathrm{curl}\ \sigma\right)_{\mu\nu} .\label{Magnetic recast}
\end{equation}
This constraint shows that the magnetic part of the Weyl tensor is induced by the curl of the shear and the coupling between the acceleration due to the fluid dynamics and the vorticity.
\end{itemize}

\subsection{Once-contracted Bianchi identities}

The second and third set of dynamical equations are contained in the Bianchi identities. The Riemann tensor satisfies the Bianchi identities as follows,
\begin{equation}
\nabla^{[\sigma\vphantom]}R^{\vphantom[\lambda\nu]}_{\ \ \ \mu\rho}=0 \
.\label{Riemann Bianchi}
\end{equation}

By substituting the splitting $\eref{Riemann tens}$ of the Riemann tensor $R^{\lambda\nu}_{\ \ \mu\rho}$ and the effective Einstein field equations $\eref{Einstein ef eq}$ into the Bianchi identities $\eref{Riemann Bianchi}$ and contracting two indices ($\sigma$ and $\rho$), the once-contracted Bianchi identities are found to be,
\begin{equation}
\nabla^{\rho}C^{\lambda\nu}_{\ \ \ \mu\rho} +
\nabla^{[\lambda\vphantom]}R^{\vphantom[\nu]}_{\ \ \mu} +
{\textstyle\frac{1}{6}}\delta_{\mu}^{\ \
[\lambda\vphantom]}\nabla^{\vphantom[\nu]}\mathcal{R}=0 \
.\label{Simple Bianchi}
\end{equation}
In a similar manner to the Ricci identities, the information stored in the once-contracted Bianchi identities has to be projected along the worldlines $u^{\mu}$ and on the orthogonal hypersurfaces $h^{\mu}_{\ \nu}$. The projections yield respectively two propagation and two constraint equations,
\begin{eqnarray}
h_{\alpha\langle\mu}h^{\gamma}_{\
\nu\rangle}u_{\beta}\left(\nabla^{\rho}C^{\alpha\beta}_{\ \ \ \gamma\rho} +
\nabla^{[\alpha\vphantom]}R^{\vphantom[\beta]}_{\ \ \gamma} +
{\textstyle\frac{1}{6}}\delta_{\gamma}^{\ \
[\alpha\vphantom]}\nabla^{\vphantom[\beta]}\mathcal{R}\right) =\,
0 \ ,\label{Bianchi prop E}\\
\eta_{\lambda\sigma\langle\mu}h^{\gamma}_{\ \nu\rangle}h^{\lambda}_{\ \alpha}h^{\sigma}_{\
\beta}\left(\nabla^{\rho}C^{\alpha\beta}_{\ \ \
\gamma\rho} + \nabla^{[\alpha\vphantom]}R^{\vphantom[\beta]}_{\ \
\gamma} + {\textstyle\frac{1}{6}}\delta_{\gamma}^{\ \
[\alpha\vphantom]}\nabla^{\vphantom[\beta]}\mathcal{R}\right) =\, 0
\
,\label{Bianchi prop H}\\
h_{\alpha\mu}h^{\gamma}_{\ \beta}\left(\nabla^{\rho}C^{\alpha\beta}_{\
\ \ \gamma\rho} + \nabla^{[\alpha\vphantom]}R^{\vphantom[\beta]}_{\
\ \gamma} + {\textstyle\frac{1}{6}}\delta_{\gamma}^{\ \
[\alpha\vphantom]}\nabla^{\vphantom[\beta]}\mathcal{R}\right) =\,
0 \ ,\label{Bianchi const E}\\
\eta_{\lambda\sigma\mu}u^{\gamma}h^{\lambda}_{\ \alpha}h^{\sigma}_{\
\beta}\left(\nabla^{\rho}C^{\alpha\beta}_{\ \ \
\gamma\rho} + \nabla^{[\alpha\vphantom]}R^{\vphantom[\beta]}_{\ \
\gamma} + {\textstyle\frac{1}{6}}\delta_{\gamma}^{\ \
[\alpha\vphantom]}\nabla^{\vphantom[\beta]}\mathcal{R}\right) =\, 0
\ .\label{Bianchi const H}
\end{eqnarray}

The sets of equations are explicitly determined by substituting the expression for the Weyl tensor splitting $\eref{Riemann decomp}$ and the Ricci tensor $\eref{Ef Ricci tensor}$ into the projections of the once-contracted Bianchi identities $\eref{Bianchi prop E}-\eref{Bianchi const H}$.

\vspace{0.5cm} The propagation equations are found to be as follows.

\begin{itemize}
\item The electric propagation equation,
\begin{eqnarray}
\eqalign{\dot{E}_{\langle\mu\nu\rangle} = &-\Theta\,E_{\mu\nu}+\left(\mathrm{curl}\
H\right)_{\mu\nu}-{\textstyle\frac{\kappa}{2}}\left(\rho_s+p_s\right)\sigma_{\mu\nu}\\
&+3{\sigma_{\langle\mu}}^{\lambda}E_{\nu\rangle\lambda}+{\omega_{\langle\mu}}^{\lambda}E_{\nu\rangle\lambda}
-2\eta_{\rho\lambda\langle\mu}{H_{\nu\rangle}}^{\lambda}a^{\rho}+\kappa\left(S_{\dot{E}}\right)_{\langle\mu\nu\rangle}\
,}\label{E prop eq}
\end{eqnarray}
where
\begin{eqnarray*}
\eqalign{\left(S_{\dot{E}}\right)_{\langle\mu\nu\rangle}=\,
&-\left({\sigma_{\langle\mu}}^{\lambda}S_{\nu\rangle\lambda}-\omega_{\langle\mu}S_{\nu\rangle}\right)^{\cdot}_{\bot}
-{\textstyle\frac{1}{3}}\,\Theta\left({\sigma_{\langle\mu}}^{\lambda}S_{\nu\rangle\lambda}-\omega_{\langle\mu}S_{\nu\rangle}\right)\\
&-{\textstyle\frac{1}{2}}\sigma_{\lambda\rho}\left({\sigma_{\langle\mu}}^{\lambda}{S_{\nu\rangle}}^{\rho}-{\omega_{\langle\mu}}^{\lambda}{S_{\nu\rangle}}^{\rho}\right)\\
&+{\textstyle\frac{1}{2}}\left(D_{\langle\mu}-2a_{\langle\mu}\right)\left(D^{\lambda}S_{\nu\rangle\lambda}-2a^{\lambda}S_{\nu\rangle\lambda}\right)
\ .}
\end{eqnarray*}
This equation is similar in form to Maxwell's electric propagation equation in an expanding universe. The
$\left(S_{\dot{E}}\right)_{\langle\mu\nu\rangle}$ term on the RHS of relation $\eref{E prop eq}$ describes how the coupling between the spin density and the fluid anisotropies contributes to the gravitational tidal field $E_{\mu\nu}$.
\item The magnetic propagation equation,
\begin{eqnarray}
\eqalign{
\dot{H}_{\langle\mu\nu\rangle} = &-
\Theta\,H_{\mu\nu}-\left(\mathrm{curl}\ E\right)_{\mu\nu}\\
&+3{\sigma_{\langle\mu}}^{\lambda}H_{\nu\rangle\lambda}-{\omega_{\langle\mu}}^{\lambda}H_{\nu\rangle\lambda}
+2\eta_{\rho\lambda\langle\mu}{E_{\nu\rangle}}^{\lambda}a^{\rho}+\kappa\left(S_{\dot{H}}\right)_{\langle\mu\nu\rangle}\
,}\label{H prop eq}
\end{eqnarray}
where
\begin{eqnarray*}
\eqalign{
\left(S_{\dot{H}}\right)_{\langle\mu\nu\rangle} =\,&{\textstyle\frac{1}{2}}\eta_{\sigma\rho\langle\mu}[D^{\sigma}\{(\sigma^{\rho\lambda}-\omega^{\rho\lambda})S_{\nu\rangle\lambda}+
S^{\rho\lambda}(\sigma_{\nu\rangle\lambda}-\omega_{\nu\rangle\lambda})\}\vphantom]
\\
&\phantom{\eta_{\sigma\rho\langle\mu}\ }-({\sigma_{\nu\rangle}}^{\sigma}-{\omega_{\nu\rangle}}^{\sigma})(D_{\lambda}S^{\rho\lambda}-2a_{\lambda}S^{\rho\lambda})\\
&\phantom{\eta_{\sigma\rho\langle\mu}\ }-\omega^{\sigma\rho}(D^{\lambda}S_{\nu\rangle\lambda}-2a^{\lambda}S_{\nu\rangle\lambda})]\
.}
\end{eqnarray*}
This expression is analogous to Maxwell's magnetic propagation equation in an expanding universe. The
$\left(S_{\dot{H}}\right)_{\langle\mu\nu\rangle}$ term on the RHS of this relation $\eref{H prop eq}$ describe how the coupling between the spin density and the fluid anisotropies contributes to the gravitational tidal field $H_{\mu\nu}$.
\end{itemize}
In a similar manner to that in which Maxwell's equations describe electrodynamics in an expanding universe, the coupling between the electric $\eref{E prop eq}$ and magnetic $\eref{H prop eq}$ propagation equations gives rise to gravitational waves damped by the expansion of the universe.

\vspace{0.5cm} The constraint equations are given by the following relations.

\begin{itemize}
\item The electric constraint equation,
\begin{equation}
D^{\lambda}E_{\mu\lambda} =
{\textstyle\frac{\kappa}{3}}D_{\mu}\rho_s-3\omega^{\lambda}H_{\mu\lambda}
-\eta_{\mu\nu\lambda}\sigma^{\nu}_{\
\rho}H^{\lambda\rho}+\kappa\left(S_{\mathrm{div}E}\right)_{\mu}
\ ,\label{E const eq}
\end{equation}
where
\begin{eqnarray*}
\eqalign{
\left(S_{\mathrm{div}E}\right)_{\mu} =
\,&D_{\mu}[(\sigma^{\lambda\rho}-\omega^{\lambda\rho})S_{\lambda\rho}]
-D^{\lambda}[(\sigma_{(\mu\vphantom)}^{\ \ \rho}-\omega_{(\mu\vphantom)}^{\ \ \rho})S_{\vphantom(\lambda)\rho}]\\
&-{\textstyle\frac{1}{3}}\Theta\,\left(D^{\lambda}S_{\mu\lambda}-2a^{\lambda}S_{\mu\lambda}\right)
+{\textstyle\frac{1}{2}}\sigma_{\mu\rho}\left(D_{\lambda}S^{\rho\lambda}-2a_{\lambda}S^{\rho\lambda}\right)\ .}
\end{eqnarray*}
This constraint is a vector analogue of the Newtonian Poisson equation. It is similar in form to Maxwell's electric divergence equation. For this gravitational field equation, the source is not
the electric charge density but the energy density. The $\left(S_{\mathrm{div}E}\right)_{\mu}$ term on the RHS of expression $\eref{E const eq}$ describes how the coupling between the spin density and the fluid anisotropies acts like an effective electric divergence source.

\item The magnetic constraint equation,
\begin{equation}
D^{\lambda}H_{\mu\lambda} =
\kappa\left(\rho_s+p_s\right)\omega_{\mu}+3\omega^{\lambda}E_{\mu\lambda}
+\eta_{\mu\nu\lambda}\sigma^{\nu}_{\
\rho}E^{\lambda\rho}+\kappa\left(S_{\mathrm{div}H}\right)_{\mu}
\ ,\label{H const eq}
\end{equation}
where
\begin{equation*}
\left(S_{\mathrm{div}H}\right)_{\mu} =
\,{\textstyle\frac{1}{2}}\eta_{\mu\nu\rho}D^{\nu}\left(D_{\lambda}S^{\rho\lambda}-2a_{\lambda}S^{\rho\lambda}\right)\ .
\end{equation*}
This constraint is analogous to Maxwell's magnetic divergence equation. Unlike for Maxwell's equation, this gravitational field equation has a source term which is the fluid vorticity. The $\left(S_{\mathrm{div}H}\right)_{\mu}$ term on the RHS of expression $\eref{H const eq}$ describes how the coupling between the spin density and the fluid anisotropies acts like an effective magnetic divergence source.
\end{itemize}

\subsection{Twice-contracted Bianchi identities}

The third set of equations is given by the twice-contracted Bianchi identities which represent the conservation of the effective stress energy momentum tensor. They are obtained by performing a second contraction ($\mu=\nu$) on the once-contracted Bianchi identities $\eref{Simple Bianchi}$,
\begin{equation}
\nabla^{\mu}\left(R_{\mu\nu}+{\textstyle\frac{1}{2}}g_{\mu\nu}\mathcal{R}\right)=
\kappa\nabla^{\mu}T^{s}_{\mu\nu}=0\ .\label{Twice Bianchi}
\end{equation}

There are only two possible projections to extract the information stored in the twice-contracted Bianchi identities,
\begin{eqnarray}
u^{\mu}\nabla^{\nu}T^{s}_{\mu\nu}=0\ ,\label{Bianchi cons En}\\
{h_{\mu}}^{\lambda}\nabla^{\nu}T^{s}_{\lambda\nu}=0\ .\label{Bianchi
cons Mom}
\end{eqnarray}

The propagation and constraint equations are explicitly determined by substituting the reduced expression for the stress-energy momentum tensor $\eref{stress energy mom 1+3}$ into the two projections of the twice-contracted Bianchi identities $\eref{Bianchi cons En}$ and $\eref{Bianchi cons Mom}$ respectively.

\vspace{0.5cm}The propagation equation is found to be as follows.

\begin{itemize}
\item The effective energy conservation equation,
\begin{equation}
\dot{\rho_s} = -\Theta\,\left(\rho_s+p_s\right)\ .\label{Eff En cons
eq}
\end{equation}
Note that for a vanishing spin density this relation reduces to the well-known energy conservation equation determining the evolution of the physical energy density $\rho$ and pressure $p$.
\end{itemize}

The constraint equation is given by the following relation.
\begin{itemize}
\item The momentum conservation equation,
\begin{equation}
D_{\mu}p_s=\left(\rho_s+p_s\right)a_{\mu}+\left(S_p\right)_{\mu}\
,\label{Mom cons eq}
\end{equation}
where
\begin{eqnarray*}
\eqalign{
\left(S_p\right)_{\mu} =
&-2\left(D^{\nu}-a^{\nu}\right)\left({\sigma_{(\mu\vphantom)}}^{\lambda}S_{\vphantom(\nu)\lambda}-{\omega_{(\mu\vphantom)}}^{\lambda}S_{\vphantom(\nu)\lambda}\right)
-\left(D^{\lambda}S_{\mu\lambda}-2a^{\lambda}S_{\mu\lambda}\right)^{\cdot}_{\bot}\\
&-{\textstyle\frac{4}{3}}\Theta\left(D^{\lambda}S_{\mu\lambda}-2a^{\lambda}S_{\mu\lambda}\right)
-\left({\sigma_{\mu}}^{\nu}-{\omega_{\mu}}^{\nu}\right)\left(D^{\lambda}S_{\nu\lambda}-2a^{\lambda}S_{\nu\lambda}\right)\ .}
\end{eqnarray*}
The term $\left(S_p\right)_{\mu}$ describes how the coupling between the spin density and the fluid anisotropies contributes to the total angular momentum.
\end{itemize}

\subsection{Spin dynamics}

The last dynamical equation for the evolution of the Weyssenhoff fluid is the spin field equation $\eref{Ef spin field equations}$. To extract the spin propagation equation, the field equation has to be twice projected on the hypersurface orthogonal to the worldline. By duality, we can write it in terms of the spin density pseudovector $S^{\mu}$ without loss of information
$\eref{Pseudovector spin}$, and we obtain:
\begin{itemize}
\item The spin propagation equation,
\begin{equation}
\dot{S}_{\langle\mu\rangle}=-\Theta\,S_{\mu}\ .\label{Spin cons vector eq}
\end{equation}
This expression $\eref{Spin cons vector eq}$ can be recast in terms of the spin-density scalar $S^2$ $\eref{Spin density
scalar}$ defined as,
\begin{equation}
S^2=-S_{\mu}S^{\mu}\ .\label{Spin density pseudo}
\end{equation}
It is then simply given by,
\begin{equation}
\dot{S}=-\Theta\,S\ .\label{Spin cons eq}
\end{equation}
This relation shows that the evolution of the spin density is the same on all scales because it is entirely determined by the volume rate of expansion of the fluid. For consistency, note that this expression implies that the spin density is inversely proportional to the volume of the fluid.

The effective energy conservation equation $\eref{Eff En cons eq}$ can now be recast in terms of the true (i.e. not effective) energy density and pressure of the fluid by substituting the spin propagation equation $\eref{Spin cons eq}$,
\begin{equation}
\dot{\rho} = -\Theta\,\left(\rho+p\right)\ .\label{En cons eq}
\end{equation}
\end{itemize}

The effective energy density $\rho_s$ and pressure $p_s$ contain spin density squared $S^2$ correction terms $\eref{Ef energy pressure}$. Thus, the spin propagation equation $\eref{Spin cons eq}$ and the energy conservation equation $\eref{En cons eq}$ imply that the spin density will rule entirely the dynamics of the fluid at early times ($\kappa S^2\gg \rho, p$), whereas, at late times, the spin contribution can safely be neglected ($\kappa S^2\ll \rho, p$).

In a cosmological context, the spin dominated era might lead to an inflationnary behaviour. This promising prospect will be analysed in detail in further work. Given that the matter dominated era is not affected by the spin contribution, the cosmological model thus reduces to the dynamical behaviour of a perfect fluid in GR. Hence, the spin density contribution from the Weyssenhoff fluid is expected to affect significantly the early time evolution of the fluid leaving the late time dynamics unchanged. Therefore, it is not currently promising as a candidate to describe dark energy.

\section{Consistency of the dynamics for an irrotational Weyssenhoff fluid with no peculiar acceleration}
\label{Section 5}

The consistency of the propagation and constraint equations can be verified by evolving the constraints. This is a tedious but straightforward task. To make the problem tractable, we chose to restrict our attention to the class of models for which the fluid dynamics is described by an irrotational flow (i.e. $\omega_{\mu\nu}=0$) with no peculiar acceleration (i.e. $a^{\mu}=0$). This ensures a hypersurface-orthogonal flow and the existence of a globally defined cosmic time. If the flow is initially irrotational, it will remain so at later times $\cite{Ellis:1998}$. 

For each space-time slicing, we can now define the curvature tensors entirely in terms of the spatial hypersurface orthogonal to the worldline. For this purpose, let us define a vector $v^{\lambda}$, which is orthogonal to the worldline, and an expansion tensor $\Theta_{\mu\nu}$ according to
\begin{equation}
v^{\lambda}u_{\lambda}=0\ ,\ \ \ \ \
\Theta_{\mu\nu}={\textstyle\frac{1}{3}}\Theta\,h_{\mu\nu}+\sigma_{\mu\nu}\ .
\end{equation}
The Ricci identities on the 3-space orthogonal to the worldline can
be defined as
\begin{equation}
2D_{[\mu\vphantom]}D_{\vphantom[\nu]}v_{\rho}={\vphantom{a}^{*}R_{\mu\nu\rho}}^{\lambda}v_{\lambda}\
,\label{Projected Ricci}
\end{equation}
where the 3-space Riemann tensor
${\vphantom{a}^{*}R_{\mu\nu\rho\lambda}}$ is related to the Riemann
tensor ${R_{\mu\nu\rho\lambda}}$ by
\begin{equation}
{\vphantom{a}^{*}R_{\rho\mu\nu\lambda}}=
{h^{\alpha}}_{\rho}{h^{\beta}}_{\mu}{h^{\gamma}}_{\nu}{h^{\delta}}_{\lambda}R_{\alpha\beta\gamma\delta}
+\Theta_{\rho\nu}\Theta_{\mu\lambda}-\Theta_{\rho\lambda}\Theta_{\mu\nu}\
.\label{Riemann tensor relations}
\end{equation}
The 3-space Ricci tensor and scalar can be obtained by contracting
the 3-space Riemann tensor with the induced 3-space metric $h_{\mu\nu}$,
\begin{eqnarray}
{\vphantom{a}^{*}R_{\mu\nu}}&=h^{\rho\lambda}{\vphantom{a}^{*}R_{\rho\mu\lambda\nu}}\
,\label{3 Ricci Tensor}\\
{\vphantom{a}^{*}\mathcal{R}}&=h^{\mu\nu}h^{\rho\lambda}{\vphantom{a}^{*}R_{\rho\mu\lambda\nu}}\
.\label{3 Ricci Scalar}
\end{eqnarray}
Using $\eref{kinematics}$, $\eref{Riemann decomp}$, $\eref{Shear
prop eq}$ and $\eref{Riemann tensor relations}$, these 3-space
curvature quantities can be recast respectively as,
\begin{eqnarray}
\eqalign{
{\vphantom{a}^{*}R^{\rho\mu}_{\ \ \ \nu\lambda}}= &-{\textstyle\frac{2}{3}}\kappa{h^{\rho}}_{[\nu\vphantom]}{h^{\mu}}_{\vphantom[\lambda]}
\rho_s-4{h^{\rho}}_{[\nu\vphantom]}{E^{\mu}}_{\vphantom[\lambda]}+2{\Theta^{\rho}}_{[\nu\vphantom]}{\Theta^{\mu}}_{\vphantom[\lambda]}\\ &-2\kappa{h^{[\rho\vphantom]}}_{[\nu\vphantom]}\left(\Theta_{\vphantom[\lambda]\sigma}S^{\vphantom[\mu]\sigma}+\Theta^{\vphantom[\mu]\sigma}S_{\vphantom[\lambda]\sigma}\right)
\ ,}\label{3 Riemann tensor}\\
{\vphantom{a}^{*}R_{\mu\nu}}=\dot{\sigma}_{\langle\mu\nu\rangle}+\Theta\,\sigma_{\mu\nu}-\kappa{\sigma_{\langle\mu}}^{\lambda}S_{\nu\rangle\lambda}
-{\textstyle\frac{1}{3}}h_{\mu\nu}\left(2\kappa\rho_s-{\textstyle\frac{2}{3}}\Theta^2+2\sigma^2\right)\
,\label{Ricci tensor eq}\\
\vphantom{a}^{*}\mathcal{R}={\textstyle\frac{2}{3}}\Theta^2-2\kappa\rho_s-2\sigma^2\
,\label{Generalised Friedmann eq}
\end{eqnarray}
where the last relation is the generalised Friedmann equation expressed in terms of the spatial curvature $\vphantom{a}^{*}\mathcal{R}$.

\subsection{Evolution of the constraints}

To determine the time evolution of the constraint equations, we shall follow Maartens' approach $\cite{Maartens:1997}$ and generalise his results to the include the presence of spin. For an irrotational Weyssenhoff fluid in absence of any peculiar acceleration, the propagation equations $\eref{Raychaudhuri eq}$, $\eref{Shear prop eq}$, $\eref{E prop eq}$, $\eref{H prop eq}$, $\eref{Spin cons vector eq}$ and $\eref{En cons eq}$, denoted by $\mathcal{P}^A=0$ where $A=\mathbf{0,\dots,5}$\ , reduce to
\begin{eqnarray}
\mathcal{P}^\mathbf{0}_{\ \mu}=\dot{S}_{\langle\mu\rangle}+\Theta S_{\mu}\ ,\label{P0}\\
\mathcal{P}^\mathbf{1}=\dot{\rho}+\Theta\left(\rho+p\right)\ ,\label{P1}\\		\mathcal{P}^\mathbf{2}=\dot{\Theta}+{\textstyle\frac{1}{3}}\Theta^2+2\sigma^2+{\textstyle\frac{\kappa}{2}}\left(\rho_s + 3p_s\right)\ ,\label{P2}\\
\mathcal{P}^\mathbf{3}_{\ \mu\nu}=\dot{\sigma}_{\langle\mu\nu\rangle}+{\textstyle\frac{2}{3}}\Theta\,\sigma_{\mu\nu}
+\sigma_{\langle\mu}^{\ \ \lambda}\sigma_{\nu\rangle\lambda}+E_{\mu\nu}-\kappa\sigma_{\langle\mu}^{\ \ \lambda}S_{\nu\rangle\lambda}\ ,\label{P3}\\
\eqalign{
\mathcal{P}^\mathbf{4}_{\ \mu\nu}=&\dot{E}_{\langle\mu\nu\rangle}+
\Theta\,E_{\mu\nu}-\left(\mathrm{curl}\
H\right)_{\mu\nu}+{\textstyle\frac{\kappa}{2}}\left(\rho_s+p_s\right)\sigma_{\mu\nu}
-3{\sigma_{\langle\mu}}^{\lambda}E_{\nu\rangle\lambda}\\
&+\kappa\left({\sigma_{\langle\mu}}^{\lambda}S_{\nu\rangle\lambda}\right)^{\cdot}_{\bot}+{\textstyle\frac{\kappa}{3}}\,\Theta{\sigma_{\langle\mu}}^{\lambda}S_{\nu\rangle\lambda}
+{\textstyle\frac{\kappa}{2}}\sigma_{\lambda\rho}{\sigma_{\langle\mu}}^{\lambda}{S_{\nu\rangle}}^{\rho}
-{\textstyle\frac{\kappa}{2}}D_{\langle\mu}D^{\lambda}S_{\nu\rangle\lambda}\ ,}\label{P4}\\
\eqalign{\mathcal{P}^\mathbf{5}_{\ \mu\nu}=&\dot{H}_{\langle\mu\nu\rangle}+\Theta\,H_{\mu\nu}+\left(\mathrm{curl}\ E\right)_{\mu\nu}-3{\sigma_{\langle\mu}}^{\lambda}H_{\nu\rangle\lambda}\\
&-{\textstyle\frac{\kappa}{2}}\eta_{\sigma\rho\langle\mu}D^{\sigma}\left(\sigma^{\rho\lambda}S_{\nu\rangle\lambda}+\sigma_{\nu\rangle\lambda}S^{\rho\lambda}\right)+{\textstyle\frac{\kappa}{2}}\eta_{\sigma\rho\langle\mu}{\sigma_{\nu\rangle}}^{\sigma}D_{\lambda}S^{\rho\lambda}\ ,}\label{P5}
\end{eqnarray}
and the constraint equations $\eref{Shear div recast}$, $\eref{Magnetic recast}$, $\eref{E const eq}$, $\eref{H const eq}$, and $\eref{Mom cons eq}$, denoted by $\mathcal{C}^A=0$ where $A=\mathbf{0,\dots,4}$\ , become
\begin{eqnarray}
\eqalign{\mathcal{C}^\mathbf{0}_{\ \mu}=&D_{\mu}p_s+2D^{\lambda}\left({\sigma_{(\mu\vphantom)}}^{\rho}S_{\vphantom(\lambda)\rho}\right)-\sigma_{\rho}^{\ \lambda}D^{\rho}S_{\mu\lambda}+\sigma_{\mu}^{\ \lambda}D^{\rho}S_{\lambda\rho}\\
&-S_{\mu}^{\ \lambda}D_{\lambda}\Theta-{\textstyle\frac{1}{2}}S_{\mu}^{\ \lambda}D^{\rho}S_{\lambda\rho}+\eta_{\mu\nu\lambda}{S^{\nu}}_{\rho}H^{\lambda\rho}\ ,}\label{C0}\\
\mathcal{C}^\mathbf{1}_{\ \mu}=D^{\lambda}\sigma_{\mu\lambda}+\kappa D^{\lambda}S_{\mu\lambda}-{\textstyle\frac{2}{3}}D_{\mu}\Theta\ ,\label{C1}\\
\mathcal{C}^\mathbf{2}_{\ \mu\nu}=\left(\mathrm{curl}\ \sigma\right)_{\mu\nu}-H_{\mu\nu}\ ,\label{C2}\\
\eqalign{\mathcal{C}^\mathbf{3}_{\ \mu}=&D^{\lambda}E_{\mu\lambda}-{\textstyle\frac{\kappa}{3}}D_{\mu}\rho_s
+\eta_{\mu\nu\lambda}\sigma^{\nu}_{\
	\rho}H^{\lambda\rho}\\ 
&+\kappa D^{\lambda}\left({\sigma_{(\mu\vphantom)}}^{\rho}S_{\vphantom(\lambda)\rho}\right)+{\textstyle\frac{\kappa}{3}}\Theta D^{\lambda}S_{\mu\lambda}-{\textstyle\frac{\kappa}{2}}\sigma_{\mu\rho}D_{\lambda}S^{\rho\lambda}\ ,}\label{C3}\\
\mathcal{C}^\mathbf{4}_{\ \mu}=D^{\lambda}H_{\mu\lambda}-\eta_{\mu\nu\lambda}\sigma^{\nu}_{\
 \rho}E^{\lambda\rho}-{\textstyle\frac{\kappa}{2}}\eta_{\mu\nu\lambda}D^{\nu}D_{\rho}S^{\lambda\rho}\ .\label{C4}
\end{eqnarray}

The evolution of the constraints $\mathcal{C}^{A}$ along the worldlines $u^{\mu}$ leads to a system of equations $\dot{\mathcal{C}}^{A}=\mathcal{F}^{A}(\mathcal{C}^{B})$, where $\mathcal{F}^{A}$ do not contain time derivatives, since these are eliminated via the propagation equations $\mathcal{P}^{A}$ and suitable identities. The covariant analysis of propagation and constraint equations involves frequent use of a number of algebraic and differential identities governing the kinematical and dynamical quantities. In particular, one requires commutation rules for spatial and time derivatives. The necessary identities are collected for convenience in {\it Appendix} C. After lengthy calculations the explicit time evolution of the constraints $\eref{C1}$, $\eref{C2}$, and $\eref{C4}$ is found to be,
\begin{eqnarray}
(\mathcal{C}^\mathbf{1}_{\ \mu})^{\cdot}_{\bot}=-\Theta\mathcal{C}^\mathbf{1}_{\ \mu}-2{\eta_{\mu}}^{\rho\sigma}{\sigma_{\sigma}}^{\lambda}\mathcal{C}^\mathbf{2}_{\ \lambda\rho}-\mathcal{C}^\mathbf{3}_{\ \mu}+\kappa\mathcal{C}^\mathbf{0}_{\ \mu}\ ,\label{tC1}\\
(\mathcal{C}^\mathbf{2}_{\ \mu\nu})^{\cdot}_{\bot}=-\Theta\mathcal{C}^\mathbf{2}_{\ \mu\nu}+{\eta^{\lambda\rho}}_{(\mu\vphantom)}\sigma_{\vphantom(\nu)\rho}\mathcal{C}^\mathbf{1}_{\ \lambda}\ ,\label{tC2}\\
\eqalign{
(\mathcal{C}^\mathbf{4}_{\ \mu})^{\cdot}_{\bot}=&-{\textstyle\frac{4}{3}}\Theta\mathcal{C}^\mathbf{4}_{\ \mu}+{\textstyle\frac{1}{2}}{\sigma_{\mu}}^{\lambda}\mathcal{C}^\mathbf{4}_{\ \lambda}+{\textstyle\frac{3}{2}}{H_{\mu}}^{\lambda}\mathcal{C}^\mathbf{1}_{\ \lambda}\\ 
&+{\eta_{\mu\rho}}^{\sigma\rho}{H_{\sigma}}^{\lambda}\mathcal{C}^\mathbf{2}_{\ \rho\lambda}-{\textstyle\frac{1}{2}}\mathrm{curl}\ \mathcal{C}^\mathbf{3}_{\ \mu}\ .}\label{tC4}
\end{eqnarray}

The constraints are preserved under evolution as now briefly explain. Suppose that the constraints are satisfied on an initial spatial hypersurface\ $\{t=t_0\}$, i.e. $\mathcal{C}^{A}|_{t_0}=0$, where $t$ is the proper time along the worldlines. Since $\mathcal{C}^{A}=0$ is a solution for the initial data, it then follows from $\eref{tC1}-\eref{tC4}$ that the constraints are satisfied for all time.

The time evolution of $\mathcal{C}^\mathbf{0}_{\ \mu}$ was not explicitly established because the equation of state needs to be specified for this endeavour. Neither was the expression for the time evolution of $\mathcal{C}^\mathbf{3}_{\ \mu}$ explicitly determined due to the overwhelming algebraic complexity of that particular computation. However, it is plausible that the dynamics is consistent since the three time evolution equations for the constraints $\eref{tC1}$, $\eref{tC2}$ and $\eref{tC4}$ involve all the constraint and propagation equations. This is true with the exception of $\mathcal{P}^\mathbf{1}$. As we discuss in detail below, $\mathcal{P}^\mathbf{1}$ is not involved in the time evolution of $\eref{tC1}$, $\eref{tC2}$ and $\eref{tC4}$. However, Obukhov and Korotky have shown $\cite{Obukhov:1987}$, using the Frenkel condition, that any perfect fluid with spin in the EC theory has an energy conservation equation of the form $\mathcal{P}^\mathbf{1}$. This is sufficient to show independently the consistency of $\mathcal{P}^\mathbf{1}$.

The time evolution of $\mathcal{C}^\mathbf{1}_{\ \mu}$, $\eref{tC1}$, involves the propagation equations $\mathcal{P}^\mathbf{0}_{\ \mu}$, $\mathcal{P}^\mathbf{2}$, $\mathcal{P}^\mathbf{3}_{\ \mu\nu}$ and the constraint equations $\mathcal{C}^\mathbf{0}_{\ \mu}$, $\mathcal{C}^\mathbf{1}_{\ \mu}$, $\mathcal{C}^\mathbf{2}_{\ \mu\nu}$. It has been determined by using the covariant identities 
$\eref{f2}$ and $\eref{A2}$.

The time evolution of $\mathcal{C}^\mathbf{2}_{\ \mu\nu}$, $\eref{tC2}$, involves the propagation equations $\mathcal{P}^\mathbf{3}_{\ \mu\nu}$, $\mathcal{P}^\mathbf{5}_{\ \mu\nu}$ and the constraint equations $\mathcal{C}^\mathbf{1}_{\ \mu}$, $\mathcal{C}^\mathbf{2}_{\ \mu\nu}$. It has been determined by using the covariant identities $\eref{A1}$ and $\eref{B3}$.

The time evolution of $\mathcal{C}^\mathbf{4}_{\ \mu}$, $\eref{tC4}$, involves the propagation equations $\mathcal{P}^\mathbf{0}_{\ \mu}$, $\mathcal{P}^\mathbf{3}_{\ \mu\nu}$, $\mathcal{P}^\mathbf{4}_{\ \mu\nu}$, $\mathcal{P}^\mathbf{5}_{\ \mu\nu}$ and the constraint equations $\mathcal{C}^\mathbf{1}_{\ \mu}$, $\mathcal{C}^\mathbf{2}_{\ \mu\nu}$, $\mathcal{C}^\mathbf{3}_{\ \mu}$, $\mathcal{C}^\mathbf{4}_{\ \mu}$. It has been determined by using the covariant identities 
$\eref{f1}$, $\eref{V1}$, $\eref{A2}$, $\eref{B1}$ and $\eref{B2}$.

The constraint equations are not linearly independent given that they satisfy, 
\begin{eqnarray}
\mathcal{C}^\mathbf{4}_{\ \mu}=-{\textstyle\frac{1}{2}}\mathrm{curl}\ \mathcal{C}^\mathbf{1}_{\ \mu}-D^{\lambda}\mathcal{C}^\mathbf{2}_{\ \mu\lambda}\ .\label{relation constraints}
\end{eqnarray}

The consistency of the constraint equations can be explicitly inferred from relation $\eref{relation constraints}$ as  explained below $\cite{Maartens:1997}$. For any given spatial hypersurface, i.e.\ $\{t=\mathrm{const}\}$, the linear dependence $\eref{relation constraints}$ of the constraint equations implies the constraint $\mathcal{C}^\mathbf{4}_{\ \mu}$ is satisfied provided that the constraints $\mathcal{C}^\mathbf{1}_{\ \mu}$ and $\mathcal{C}^\mathbf{2}_{\ \mu\nu}$ are also satisfied. Moreover, the time evolution of $\mathcal{C}^\mathbf{1}_{\ \mu}$ and $\mathcal{C}^\mathbf{2}_{\ \mu\nu}$, described by $\eref{tC1}$ and $\eref{tC2}$ respectively, depends explicitly on $\mathcal{C}^\mathbf{0}_{\ \mu}$ and $\mathcal{C}^\mathbf{3}_{\ \mu}$. Hence, if we take $\mathcal{C}^\mathbf{0}_{\ \mu}$ as determining $D^{\lambda}S_{\mu\lambda}$, $\mathcal{C}^\mathbf{1}_{\ \mu}$ as defining $D_{\mu}\Theta$, $\mathcal{C}^\mathbf{2}_{\ \mu\nu}$ as establishing $H_{\mu\nu}$ and $\mathcal{C}^\mathbf{3}_{\ \mu}$ as setting $D_{\mu}\rho_s$, the constraint equations are consistent with each other because $\mathcal{C}^\mathbf{4}_{\ \mu}$ then follows.
 
The consistency of the constraints for a perfect fluid in GR with non vanishing vorticity and peculiar acceleration has been established by van Elst $\cite{vanElst:1996}$. Thus, having shown that the dynamics of an irrotational Weyssenhoff fluid in absence of any peculiar acceleration ($\omega=a=0$) is consistent, it is very plausible $\--$ although not proven $\--$ that this will remain the case in the general case when the vorticity and the peculiar acceleration are considered. Hence, in that case, to establish explicitly the consistency of the constraints for such a fluid, the coherence of the terms involving the coupling between the spin density, the vorticity density and the peculiar acceleration would have to be shown respectively. This would be a extremely laborious algebraic task, but it is, in fact, quite likely to be true since the consistency of two different particular cases has already been established.  

\section{Comparison with previous results}

A first attempt to study the dynamics of a Weyssenhoff fluid in a $1+3$ covariant approach was initiated by Palle $\cite{Palle:1998}$. The results we find in this paper disagree, however, with the majority of the results derived by Palle, as we now briefly explain.

In a similar way to our own procedure, Palle based his analysis on the effective Einstein field equations for a Weyssenhoff fluid obtained by Obukhov and Korotky $\cite{Obukhov:1987}$, which are outlined in relation $\{1\}$ of his publication. As explicitly stated in his work, Palle projects the EC version of the Ricci identities determined by Hehl $\cite{Hehl:1974}$,
\begin{equation}
2\tilde{\nabla}_{[\mu\vphantom]}\tilde{\nabla}_{\vphantom[\nu]}u_{\rho}
={\tilde{R}_{\mu\nu\rho}}^{\ \ \ \ \lambda}u_{\lambda}
+2{Q^{\lambda}}_{\mu\nu}\tilde{\nabla}_{\lambda}u_{\rho},
\label{Ricci Hehl}
\end{equation}
which are given in relation $\{4\}$ of his paper to find the corresponding propagation and constraint equations. This stands in direct contradiction with the fact that the $1+3$ covariant approach used is based on effective GR field equations.

Moreover, in Palle's work, there is no mention of the antisymmetric part of the EC field equations which lead to the spin field equation. It seems unfeasible to provide an accurate description of a cosmological fluid with spin without describing the spin dynamics.

Furthermore, Palle chose to neglect the contributions due to the electric and magnetic part of the Weyl tensor but did not provide any explanation for this. Indeed, the relation $\{7\}$ he obtained for the shear propagation equation has no tidal gravitational field $E_{\mu\nu}$ contribution, and there is no magnetic constraint equation. To describe the late time cosmological evolution, it seems indeed reasonable to neglect the contributions due to the primordial free propagating gravitational fields which
have been damped by the cosmological expansion. However, these fields do significantly affect the early dynamics and have to be taken into account in a general description of cosmological models.

Finally, Palle does not determine the cosmological relations derived from the Bianchi identities. Again, these would be very useful to understand the dynamics of the early time evolution of cosmological models.

Palle has recently clarified $\cite{Palle:2007}$ certain points relating to the approach he followed in analysing the cosmological implications of a Weyssenhoff fluid. Several issues still, however, remain a concern, as outlined below.
 
It is perfectly legitimate to analyse the Weyssenhoff fluid dynamics within an EC framework without resorting to an effective GR framework. In such a case, the appropriate way to determine the large scale propagation and constraint equations is indeed to project the EC Ricci identites $\eref{Ricci Hehl}$ on the relevant hypersurfaces, which is what Palle seems to have done. To achieve this, the EC Ricci identities have to be explicitly determined using the effective EC field equations. Our contention is that the only effective field equations \{1\} mentioned in Palle's paper $\cite{Palle:1998}$, and used to perform the calculations, are the effective GR field equations obtained by Obukhov and Korotky. We believe that the GR field equations are incompatible with EC Ricci identities, which would thus invalidate the analysis.

The physical motivation for using GR field equations is that it provides a more natural generalisation for the dynamics of a perfect fluid within GR. Although Palle's procedure seems inconsistent, we have nevertheless translated his results within a GR framework to be able to compare them. To compare explicitly our results with those obtained by Palle, note that the torsion scalar $Q$ he uses is related $\--$ due to the algebraic coupling between spin and torsion $\--$ to our definition of the spin density $S$ by,
\begin{equation}
Q=\kappa S\ .
\end{equation}
It is now straightforward to see that neither the propagation equations $\{5-7\}$ nor the constraint equations $\{8-9\}$ he found agree with our own corresponding results. The detailed comparision and analysis can be found in {\it Appendix} D. We hope that it might clarify this particular issue.
 
With regard to the scope of Palle's paper, on large scales, the contribution of the tidal forces to dynamics of the Weyssenhoff fluid can indeed be neglected. Hence, the Weyl tensor can safely be ignored in his approach, but it was not stated by Palle that only the dynamics on large scales were under consideration. It was important to clarify this issue because we have considered the dynamical evolution of Weyssenhoff fluid on all scales.
 
Finally, let us just mention that, as suggested by Palle in $\cite{Palle:2007}$, it might indeed be more appropriate to consider an N-body simulation to determine the large scale and late time dynamics of a Weyssenhoff fluid in a cosmological context, as Palle suggested. However, this seems to us to lie outside our study, as we simply considered the evolution of such a fluid on all scales and for all times.

\section{Conclusions}

We have used the $1+3$ covariant approach to determine the dynamics of a Weyssenhoff fluid in a non-perturbative and hence completely general manner. This gauge-invariant procedure leads to a consistent set of seven propagation and six constraint equations. These give respectively the time and spatial covariant derivative of the set of dynamical variables ($\rho$, $\Theta$, $\sigma$, $\omega$, $E$, $B$, $S$). Compared to the dynamics of a perfect fluid in GR, there is one additional propagation equation which is the spin density propagation equation. Note that the spin constraint is included in the shear constraint.

\ack

S~D~B thanks the Isaac Newton Studentship and the Sunburst Fund for their support. The authors also thank Anthony Challinor for insightful and stimulating discussions.

\appendix

\section{1+3 covariant formalism}
\label{Appendix A}

We will briefly outline the basics of the $1+3$ covariant formalism introduced by Hawking and extended by Ellis to describe the fluid dynamics within GR in a non-perturbative way. The aim of this approach is to study the intrinsic dynamics of fluid models in a physically transparent manner. This formalism relies on covariantly defined variables, which are gauge-invariant by construction, thus simplifying the methodology and clarifying the physical interpretation of the models. Furthermore, the form of the metric does not need to be explicitly specified and can remain fully general until the dynamics is determined. Finally, this approach admits a covariant and gauge-invariant linearization that allows linearized calculations to be performed in a direct manner $\cite{Challinor:2000}$.

To introduce the $1+3$ covariant formalism, we follow Ellis and Van Elst's approach $\cite{Ellis:1998}$ using the opposite signature. The approach is based on a $1+3$ decomposition of geometric quantities with respect to a fundamental $4$-velocity $u^{\mu}$ which uniquely determines the worldline of every infinitesimal volume element of fluid,
\begin{equation}
u^{\mu}=\frac{dx^{\mu}}{d\tau}\ ,\ \ \ \ \ \ \ u_{\mu}u^{\mu}=1\ ,
\end{equation}
where $\tau$ is the proper time measured along the worldlines. In the context of a general cosmological model, we require that the $4$-velocity be chosen in a physical manner such that in the FRW limit the dipole of the cosmic microwave background radiation vanishes. This condition is necessary to ensure the gauge-invariance of the approach.

The $4$-velocity $u^{\mu}$ defines locally two projection tensors in a unique fashion,
\begin{eqnarray}
U_{\mu\nu} = u_{\mu}u_{\nu} \ \ \ &\Rightarrow \ \ \
U^{\mu}_{\ \lambda}U^{\lambda}_{\ \nu} = U^{\mu}_{\ \nu} \ , \ \
&U^{\mu}_{\ \mu} = 1 \ , \ \ U_{\mu\nu}u^{\nu} = u_{\mu}\ ,\\
h_{\mu\nu} = g_{\mu\nu}-u_{\mu}u_{\nu} \ \ \ &\Rightarrow \ \ \
h^{\mu}_{\ \lambda}h^{\lambda}_{\ \nu} = h^{\mu}_{\ \nu} \ , \ \
&h^{\mu}_{\ \mu} = 3 \ , \ \ h_{\mu\nu}u^{\nu} = 0\ .
\end{eqnarray}

The first projects parallel to the 4-velocity vector $u^{\mu}$, and the second determines the (orthogonal) metric properties of the instantaneous rest-spaces of observers moving with $4$-velocity $u^{\mu}$. There is also a volume element for the rest-spaces defined as
\begin{equation}
\eta_{\mu\nu\rho} = u^{\lambda}\eta_{\lambda\mu\nu\rho} \ \ \ \
\Rightarrow \ \ \ \ \eta_{\mu\nu\rho} = \eta_{[\mu\nu\rho]} \ , \ \
\eta_{\mu\nu\rho}u^{\rho} = 0 \ ,
\end{equation}
where $\eta_{\lambda\mu\nu\rho}$ is the 4-dimensional volume element ($\eta_{\lambda\mu\nu\rho} = \eta_{[\lambda\mu\nu\rho]}$, $\eta_{0123} = \sqrt{|\det\,g_{\mu\nu}\,|}$). Note that the contraction of the rest-space volume elements can be expressed in terms of the induced metric on these rest-spaces as
\begin{equation}
\eta_{\alpha\beta\gamma}\eta^{\mu\nu\rho}= -3!h^{[\mu\vphantom]}_{\ \
\alpha}h^{\nu}_{\ \beta}h^{\vphantom[\rho]}_{\ \ \gamma} =
-3!h^{\mu}_{\ [\alpha\vphantom]}h^{\nu}_{\
\vphantom[\beta\vphantom]}h^{\rho}_{\ \vphantom[\gamma]}\ .
\end{equation}
Moreover, we define two projected covariant derivatives which are the time projected covariant derivative along the worldline (denoted $\mathbf{\dot{}}\ $) and the orthogonally projected covariant derivative (denoted $D_{\mu}$). For any general tensor $T^{\mu\dots}_{\ \ \ \ \nu\dots}$, these are respectively defined as
\begin{eqnarray}
\dot{T}^{\mu\dots}_{\ \ \ \ \nu\dots} &\equiv
u^{\lambda}\nabla_{\lambda}T^{\mu\dots}_{\ \ \ \ \nu\dots}\ ,\\
D_{\lambda}T^{\mu\dots}_{\ \ \ \ \nu\dots} &\equiv h^{\epsilon}_{\
\lambda}h^{\mu}_{\ \rho}\dots h^{\sigma}_{\ \nu}\dots
\nabla_{\epsilon}T^{\rho\dots}_{\ \ \ \sigma\dots}\ .
\end{eqnarray}
Furthermore, the dynamics is determined by projected tensors that are orthogonal to $u^{\mu}$ on every index. The angle brackets are used to denote respectively orthogonal projections of vectors and the orthogonally projected symmetric trace-free part $(\mathrm{PSTF})$ of rank-$2$ tensors according to,
\begin{eqnarray}
v^{\langle\mu\rangle} &= h^{\mu}_{\ \nu}v^{\nu}\ ,\\
T^{\langle \mu\nu\rangle} &= \left(h^{(\mu\vphantom)}_{\ \
\rho}h^{\vphantom(\nu)}_{\ \ \sigma}-
{\textstyle\frac{1}{3}}h^{\mu\nu}h_{\rho\sigma}\right)T^{\rho\sigma}\ .
\end{eqnarray}
For convenience, the angle brackets are also used to denote the orthogonal projections of covariant time derivatives of tensors along the worldline $u^{\mu}$ as follows,
\begin{eqnarray}
\dot{v}^{\langle\mu\rangle} &= h^{\mu}_{\ \nu}\dot{v}^{\nu}\ ,\\
\dot{T}^{\langle \mu\nu\rangle} &= \left(h^{(\mu\vphantom)}_{\ \
\rho}h^{\vphantom(\nu)}_{\ \ \sigma}-
{\textstyle\frac{1}{3}}h^{\mu\nu}h_{\rho\sigma}\right)\dot{T}^{\rho\sigma}\ .
\end{eqnarray}
The orthogonal projection of the covariant time derivative of a general tensor $T^{\mu\dots}_{\ \ \ \nu\dots}$ is denoted by,
\begin{equation}
\left(T^{\mu\dots}_{\ \ \ \nu\dots}\right)^{\cdot}_{\bot}\equiv h^{\mu}_{\ \rho}\dots h^{\sigma}_{\ \nu}\dots
u^{\lambda}\nabla_{\lambda}T^{\rho\dots}_{\ \ \ \sigma\dots}\ .
\end{equation}
It is also useful to define the projected covariant curl as,
\begin{equation}
(\mathrm{curl}\ T)_{\mu\dots\nu}\equiv\eta_{\rho\sigma\langle\mu}
D^{\rho}T_{\ \dots\nu\rangle}\phantom{}^{\sigma}\ .\label{curl}
\end{equation}

Information relating to the kinematics is contained in the covariant derivative of $u^{\mu}$ which can be split into irreducible parts, defined by their symmetry properties,
\begin{equation}
\nabla_{\mu}u_{\nu}=u_{\mu}a_{\nu}+D_{\mu}u_{\nu}=
u_{\mu}a_{\nu}+{\textstyle\frac{1}{3}}\Theta
h_{\mu\nu}+\sigma_{\mu\nu}+\omega_{\mu\nu}\ ,\label{kinematics}
\end{equation}
where
\begin{itemize}
\item $a^{\mu}\equiv u^{\nu}\nabla_{\nu}u^{\mu}$ is the relativistic peculiar acceleration vector, representing the degree to which matter moves under forces other than gravity.
\item $\Theta\equiv D_{\mu}u^{\mu}$ is the scalar describing the volume rate of expansion of the fluid (with $H={\textstyle\frac{1}{3}}\Theta$ the Hubble parameter).
\item $\sigma_{\mu\nu}\equiv D_{\langle\mu}u_{\nu\rangle}$ is the trace-free rate-of-shear tensor describing the rate of distortion of the matter flow.
\item $\omega_{\mu\nu}\equiv D_{[\mu}u_{\nu]}$ is the anti-symmetric vorticity tensor describing the rotation of matter relative to a non-rotating frame.
\end{itemize}
These kinematical quantities have the following properties,
\begin{eqnarray}
a_{\mu}u^{\mu}=0\ , \ \ &\phantom{\sigma_{\nu\mu}=\sigma_{\mu\nu}}\ \ \ &\phantom{\sigma^{\mu}_{\ \mu}=0}\\
\sigma_{\mu\nu}u^{\nu}=0\ ,\ \ &\sigma_{\nu\mu}=\sigma_{\mu\nu}\ ,\ \ &\sigma^{\mu}_{\ \mu}=0\ ,\\ \omega_{\mu\nu}u^{\nu}=0\ ,\ \ &\omega_{\nu\mu}=-\omega_{\mu\nu}\ ,\ \ &\omega^{\mu}_{\ \mu}=0\ .
\end{eqnarray}
It is useful to introduce two additional pseudovectors known respectively as the vorticity and spin density. These pseudovectors are defined by duality as,
\begin{eqnarray}
\omega^{\lambda}={\textstyle\frac{1}{2}}\eta^{\lambda\mu\nu}\omega_{\mu\nu}\ \
\ \ &\Rightarrow\ \ \ \
&\omega_{\mu\nu}=-\eta_{\mu\nu\lambda}\omega^{\lambda}\ ,\\
S^{\lambda}={\textstyle\frac{1}{2}}\eta^{\lambda\mu\nu}S_{\mu\nu}\ \ \ \
&\Rightarrow\ \ \ \ &S_{\mu\nu}=-\eta_{\mu\nu\lambda}S^{\lambda}\ ,
\label{Pseudovector spin}
\end{eqnarray}
and satisfy
\begin{eqnarray}
\omega_{\mu}u^{\mu}=0\ ,\ \ &\omega_{\mu\nu}\omega^{\nu}=0\ ,\\
S_{\mu}u^{\mu}=0\ ,\ \ &S_{\mu\nu}S^{\nu}=0\ .
\end{eqnarray}
It is also of physical interest to introduce three further scalars which are respectively the acceleration, the shear and the vorticity magnitudes defined as,
\begin{eqnarray}
a^2&={\textstyle\frac{1}{2}}a_{\mu}a^{\mu}\geq 0\ ,\\
\sigma^2&={\textstyle\frac{1}{2}}\sigma_{\mu\nu}\sigma^{\mu\nu}\geq 0\ ,\\
\omega^2&={\textstyle\frac{1}{2}}\omega_{\mu\nu}\omega^{\mu\nu}\geq 0\ .
\end{eqnarray}

\section{Transformation of physical quantities under a signature change}
\label{Appendix B}

The signature convention $(+,-,-,-)$ we have used throughout this paper is the opposite of the one $(-,+,+,+)$ adopted by many authors, such as Ellis and Hawking. To facilitate the comparison between results obtained using different conventions, the explicit transformations for physical quantities evaluated within the effective field theory are given below.

The metrics, the Levi-Civita tensors and the derivatives transform as,
\begin{eqnarray*}
g_{\mu\nu}\rightarrow-g_{\mu\nu}\ ,\ \ \ \ \ \ &h_{\mu\nu}\rightarrow-h_{\mu\nu}\ ,\ \ \ \ \ \ &\eta_{\mu\nu\lambda\rho}\rightarrow\eta_{\mu\nu\lambda\rho}\ ,\ \ \ \ \ \ \eta_{\mu\nu\lambda}\rightarrow\eta_{\mu\nu\lambda}\ ,\\
\partial_{\mu}\rightarrow\partial_{\mu}\ ,\ \ \ \ \ \ \  \ \ \ &\nabla_{\mu}\rightarrow\nabla_{\mu}\ ,\ \ \ \ \ \ \ \ \ &D_{\mu}\rightarrow D_{\mu}\ .
\end{eqnarray*}

The kinematical quantities transform as,
\begin{eqnarray*}
u^{\mu}\rightarrow u^{\mu}\ ,\ \ \ \ \ \ \ \ \ &u_{\mu}\rightarrow -u_{\mu}\ ,\ \ \ \ \ \ \ 
&a^{\mu}\rightarrow a^{\mu}\ ,\ \ \ \ \ \ \ \ a_{\mu}\rightarrow -a_{\mu}\ ,\\
\sigma_{\mu\nu}\rightarrow -\sigma_{\mu\nu}\ ,\ \ \ \ \ \ &\omega_{\mu\nu}\rightarrow -\omega_{\mu\nu}\ ,\ \ \ \ \ \ &\omega^{\mu}\rightarrow \omega^{\mu}\ ,\ \ \ \ \ \ \ \omega_{\mu}\rightarrow -\omega_{\mu}\ .
\end{eqnarray*}

The dynamical quantities transform as,
\begin{eqnarray*}
R_{\mu\nu\lambda\rho}\rightarrow -R_{\mu\nu\lambda\rho}\ ,\ \ \ \ \ \ \ \ \ \ &R_{\mu\nu}\rightarrow R_{\mu\nu}\ ,\ \ \ \ \ \ \ \ \ \ 
&R\rightarrow -R\ ,\\
C_{\mu\nu\lambda\rho}\rightarrow -C_{\mu\nu\lambda\rho}\ ,\ \ \ \ \ \ \ \  \ \ &E_{\mu\nu}\rightarrow -E_{\mu\nu}\ ,\ \ \ \ \ \ \ \ 
&H_{\mu\nu}\rightarrow -H_{\mu\nu}\ ,\\
T_{\mu\nu}\rightarrow T_{\mu\nu}\ ,\ \ \ \ \ \ \ \ \ \ \ \ \ \ \ \ \ \ &S_{\mu\nu}\rightarrow S_{\mu\nu}\ ,\ \ \ \ \ \ \ \ \ \ \ 
&S^{\mu}\rightarrow -S^{\mu}\ .
\end{eqnarray*}

It is obvious that rising or lowering indices affects the sign of the transformation for any physical quantity since the space-time metric $g_{\mu\nu}$ and the spatial metric $h_{\mu\nu}$ change sign under such a transformation.   

\section{Covariant identites for an irrotational Weyssenhoff fluid with no peculiar acceleration}
\label{Appendix C}

It is straightforward to show that the derivatives of the induced metric $h_{\mu\nu}$ and the Levi-Civita tensor $\eta_{\mu\nu\lambda}$ vanish,
\begin{eqnarray}
	D_{\rho}h_{\mu\nu}=0\ ,\ \ \ \ \ \ \ \ \ \ \ \ \ \ \ &(h_{\mu\nu})^{\cdot}_{\bot}=0\ ,\\
	D_{\rho}\eta_{\mu\nu\lambda}=0\ ,\ \ \ \ \ \ \ \ \ \ \ \ \ \ &(\eta_{\mu\nu\lambda})^{\cdot}_{\bot}=0\ .
\end{eqnarray}

In this appendix, we consider an irrotational Weyssenhoff fluid ($\omega_{\mu\nu}=0$) with no peculiar acceleration ($a^{\mu}=0$). The covariant identities are defined in terms of a scalar field $f$, a vector field $V_{\mu}$ and three tensor fields, $A_{\mu\nu}$, $B_{\mu\nu}$ and $C_{\mu\nu}$ satisfying the following properties
\begin{eqnarray*}
	V_{\mu}u^{\mu}=0\ ,\ \ \ \ \ \ \ \ \ \ \ \ \ \ \ &A_{\mu\nu}u^{\mu}=A_{\nu\mu}u^{\mu}=0\ ,\\
	B_{\mu\nu}=B_{\langle\mu\nu\rangle}\ ,\ \ \ \ \ \ \ \ \ \ \ \ &C_{\mu\nu}=C_{\langle\mu\nu\rangle}\ .
\end{eqnarray*}

Using the kinematical decomposition $\eref{kinematics}$, the identities involving the derivatives of the scalar field $f$ are found to be,
\begin{eqnarray}
	D_{[\mu\vphantom]}D_{\vphantom[\nu]}f=0\ ,\label{f1}\\
	(D_{\mu}f)^{\cdot}_{\bot}=D_{\mu}\dot{f}-{\textstyle\frac{1}{3}}\Theta D_{\mu}f-{\sigma_{\mu}}^{\lambda}D_{\lambda}f\ .\label{f2}
\end{eqnarray}

Using the Ricci identities $\eref{Ricci identities}$, the identities involving the derivatives of the vector field $V_{\mu}$ and tensor field $A_{\mu\nu}$ are given by,
\begin{eqnarray}
\eqalign{(D_{\mu}V_{\nu})^{\cdot}_{\bot}=&D_{\mu}\dot{V}_{\nu}-{\textstyle\frac{1}{3}}\Theta D_{\mu}V_{\nu}-\sigma_{\mu\lambda}D^{\lambda}V_{\nu}+\eta_{\nu\lambda\rho}V^{\lambda}{H_{\mu}}^{\rho}\\
&-\kappa h_{\mu[\nu\vphantom]}V^{\rho}D^{\lambda}S_{\vphantom[\rho]\lambda}\ ,}\label{V1}\\
(D^{\lambda}V_{\lambda})^{\cdot}_{\bot}=D^{\lambda}\dot{V}_{\lambda}-{\textstyle\frac{1}{3}}\Theta D^{\lambda}V_{\lambda}-{\sigma_{\rho\lambda}}D^{\lambda}V^{\rho}-\kappa V^{\rho}D^{\lambda}S_{\rho\lambda}\ ,\label{V2}\\
\eqalign{ (D_{\lambda}A_{\mu\nu})^{\cdot}_{\bot}=&D_{\lambda}\dot{A}_{\mu\nu}-{\textstyle\frac{1}{3}}\Theta D_{\lambda}A_{\mu\nu}-\sigma_{\lambda\rho}D^{\rho}A_{\mu\nu}\\
&+\left(\eta_{\mu\sigma\rho}{A^{\sigma}}_{\nu}+\eta_{\nu\sigma\rho}{A_{\mu}}^{\sigma}\right){H_{\lambda}}^{\rho}\\
&-\kappa\left({A^{\sigma}}_{\nu}h_{\lambda[\mu\vphantom]}+{A_{\mu}}^{\sigma}h_{\lambda[\nu\vphantom]}\right)D^{\rho}S_{\vphantom[\sigma]\rho}\ ,}\label{A1}\\
\eqalign{
(D^{\lambda}A_{\mu\lambda})^{\cdot}_{\bot}=&D^{\lambda}\dot{A}_{\mu\lambda}-{\textstyle\frac{1}{3}}\Theta D^{\lambda}A_{\mu\lambda}-\sigma_{\rho\lambda}D^{\rho}{A_{\mu}}^{\lambda}+\eta_{\mu\sigma\rho}{A^{\sigma}}_{\lambda}H^{\rho\lambda}\\
&-{\textstyle\frac{\kappa}{2}}\left({A^{\sigma}}_{\mu}+2{A_{\mu}}^{\sigma}\right)D^{\rho}S_{\sigma\rho}\ .}\label{A2}
\end{eqnarray}

Using the definition of the curl $\eref{curl}$ and the spatial Ricci identities $\eref{Projected Ricci}$,the identities involving the derivatives of the symmetric trace-free tensor fields $B_{\mu\nu}$ and $C_{\mu\nu}$ yield,
\begin{eqnarray}
\eta_{\mu\nu\rho}{C^{\nu}}_{\lambda}\left(\mathrm{curl}B\right)^{\rho\lambda}=
-2C^{\rho\lambda}D_{[\mu\vphantom]}B_{\vphantom[\rho]\lambda}+{\textstyle\frac{1}{2}}C_{\mu\rho}D_{\lambda}B^{\rho\lambda}\ ,\label{B1}\\
\eqalign{
D^{\lambda}(\mathrm{curl}B)_{\mu\lambda}=&{\textstyle\frac{1}{2}}\eta_{\mu\nu\rho}D^{\nu}\left(D_{\lambda}B^{\rho\lambda}\right)+\eta_{\mu\nu\rho}{B^{\rho}}_{\lambda}\left({\textstyle\frac{1}{3}}\Theta\sigma^{\nu\lambda}-E^{\nu\lambda}\right)\\
&+{\textstyle\frac{1}{2}}\eta_{\lambda\nu\rho}{\sigma^{\lambda}}_{\mu}{\sigma^{\nu}}_{\sigma}B^{\rho\sigma}-{\textstyle\frac{3}{2}}\kappa\eta_{\mu\nu\rho}{\sigma^{\langle\nu}}_{\lambda}S^{\sigma\rangle\lambda}{B_{\sigma}}^{\rho}\ .}\label{B2}\\
\eqalign{
(\mathrm{curl}B)^{\cdot}_{\bot\mu\nu}=&(\mathrm{curl}\dot{B})_{\mu\nu}-{\textstyle\frac{1}{3}}\Theta\left(\mathrm{curl}B\right)_{\mu\nu}-{\sigma_{\sigma}}^{\lambda}\eta_{\lambda\rho\langle\mu}D^{\sigma}{B_{\rangle\nu}}^{\rho}\\
&+3{H_{\langle\mu}}^{\lambda}B_{\nu\rangle\lambda}-{\textstyle\frac{\kappa}{2}}\eta_{\lambda\rho\langle\mu}{B_{\nu\rangle}}^{\lambda}D_{\sigma}S^{\rho\sigma}\ .}\label{B3}
\end{eqnarray}

\section{Explicit comparision with Palle's results}
\label{Appendix D}

To compare our results (BHL) explicitly with the corresponding results obtained by Palle, we reexpressed his EC propagation and constraint equations $\--$ presumably obtained within an EC framework $\--$ into a GR framework using the relations given in {\it Appendix} B. The correspondence between the EC and GR connections is given by,
\begin{eqnarray}
{\tilde{\Gamma}^{\lambda}}_{\ \mu\nu}={\Gamma^{\lambda}}_{\mu\nu}+\kappa\left(u^{\lambda}S_{\mu\nu}
+u_{\mu}{S_{\nu}}^{\lambda}+u_{\nu}{S_{\mu}}^{\lambda}\right)\ ,
\end{eqnarray}
and necessary to recast the EC covariant derivative $\tilde{\nabla}_{\mu}$ in terms of its GR counterpart  $\nabla_{\mu}$. To be consistent with Palle's procedure, we only considered the dynamics on large scales, hence neglecting the contribution due to the tidal forces ($E_{\mu\nu}=H_{\mu\nu}=0$).

The propagation equations are respectively found to be (where we highlight in bold face the terms that differ):

\begin{eqnarray}
\fl \mathrm{(Palle) : } \quad 
\dot{\Theta}=-{\textstyle\frac{1}{3}}\Theta^2+D_{\lambda}a^{\lambda}+2\left(\omega^2-\sigma^2-a^2\right)-{\textstyle\frac{\kappa}{2}}\left(\rho_s + 3p_s\right)\ ,\\	
\fl \mathrm{(BHL) : } \quad
\dot{\Theta}=-{\textstyle\frac{1}{3}}\Theta^2+D_{\lambda}a^{\lambda}+2\left(\omega^2-\sigma^2-a^2\right)-{\textstyle\frac{\kappa}{2}}\left(\rho_s + 3p_s +
\bm{8\omega^{\lambda}S_{\lambda}}\right)\ .
\end{eqnarray}

\begin{eqnarray}
\fl \mathrm{(Palle) : } \quad
\dot{\omega}_{\langle\mu\rangle}=-{\textstyle\frac{2}{3}}\Theta\,\omega_{\mu}+{\textstyle\frac{1}{2}}\left(\mathrm{curl}\ a\right)_{\mu}+\left(\sigma_{\mu}^{\ \lambda}+\bm{\kappa S_{\mu}^{\ \lambda}}\right)\omega_{\lambda}\ ,\\	
\fl \mathrm{(BHL) : } \quad	\dot{\omega}_{\langle\mu\rangle}=-{\textstyle\frac{2}{3}}\Theta\,\omega_{\mu}+{\textstyle\frac{1}{2}}\left(\mathrm{curl}\ a\right)_{\mu}+\sigma_{\mu}^{\ \lambda}\omega_{\lambda}\ .
\end{eqnarray}

\begin{eqnarray}
\fl\mathrm{(Palle) : } \quad	 \eqalign{\dot{\sigma}_{\langle\mu\nu\rangle}=&-{\textstyle\frac{2}{3}}\Theta\,\sigma_{\mu\nu}+D_{\langle\mu}a_{\nu\rangle}-a_{\langle\mu}a_{\nu\rangle}-\sigma_{\langle\mu}^{\ \ \lambda}\sigma_{\nu\rangle\lambda}+\omega_{\langle\mu}\omega_{\nu\rangle}\\
&+\bm{2\kappa\sigma_{\langle\mu}^{\ \ \lambda}S_{\nu\rangle\lambda}-{\textstyle\frac{1}{3}}h_{\mu\nu}\left(\Theta-2D_{\lambda}a^{\lambda}+2a^2+2\kappa^2S^2\right)}\ ,}\\
\fl\mathrm{(BHL) : } \quad
\eqalign{\dot{\sigma}_{\langle\mu\nu\rangle}=&-{\textstyle\frac{2}{3}}\Theta\,\sigma_{\mu\nu}+D_{\langle\mu}a_{\nu\rangle}-a_{\langle\mu}a_{\nu\rangle}
-\sigma_{\langle\mu}^{\ \ \lambda}\sigma_{\nu\rangle\lambda}+
\omega_{\langle\mu}\omega_{\nu\rangle}\\
&+\bm{\kappa\left(\sigma_{\langle\mu}^{\ \
\lambda}S_{\nu\rangle\lambda}-\omega_{\langle\mu}S_{\nu\rangle}\right)}\ .}
\end{eqnarray}

The constraint equations respectively yield (where we highlight in bold face the terms that differ):

\begin{eqnarray}
\fl\mathrm{(Palle) : } \quad D_{\lambda}\left(\omega^{\lambda}+\bm{\kappa S^{\lambda}}\right)=-a_{\lambda}\left(\omega^{\lambda}+\bm{\kappa S^{\lambda}}\right)\ ,\\
\fl\mathrm{(BHL) : }  \quad D_{\lambda}\omega^{\lambda}=-a_{\lambda}\omega^{\lambda}\ .
\end{eqnarray}

\begin{eqnarray}
\fl\mathrm{(Palle) : } \quad  D_{\lambda}\left({\sigma_{\mu}}^{\lambda}+{\omega_{\mu}}^{\lambda}\right)-{\textstyle\frac{2}{3}}D_{\mu}\Theta=2a_{\lambda}{\omega_{\mu}}^{\lambda}\ ,\\
\fl\mathrm{(BHL) : } \quad
D_{\lambda}\left({\sigma_{\mu}}^{\lambda}+{\omega_{\mu}}^{\lambda}+\bm{\kappa S_{\mu}^{\ \lambda}}\right)
-{\textstyle\frac{2}{3}}D_{\mu}\Theta=2a_{\lambda}\left({\omega_{\mu}}^{\lambda}+\bm{\kappa
{S_{\mu}}^{\lambda}}\right)\ .
\end{eqnarray}

Note that the shear propagation equation is by definition trace free. This result is recovered by BHL but not by Palle. Furthermore, in absence of torsion $\--$ i.e. for a vanishing spin contribution $\--$ the shear evolution equation obtained by Palle does not reduce to Hawking and Ellis' result whereas the relation obtained by BHL does.

We could not rigorously verify Palle's result by evolving the constraints because the spin contribution to the Bianchi identities are needed for that purpose as shown in $\Sref{Section 5}$. However, it would be of considerable interest if Palle could emulate BHL and demonstrate that his set of equations also reduce to Hawking and Ellis' results in absence of torsion, and that the consistency of his equations could be established in the absence of voritcity and of any peculiar acceleration.


\section*{References}

\begin{thebibliography}{}

\bibitem{Cartan:1922}
Cartan E 1922 Sur une g\'en\'eralisation de la notion de courbure de Riemann et les espaces \`a torsion {\it Comptes Rendus de l'Acad\'emie des Sciences} {\bf 174} 593

\bibitem{Hehl:1973}
Hehl F W 1973 Spin and torsion in general relativity: I. Foundations {\it Gen. Rel. Grav.} {\bf 4} 333

\bibitem{Hehl:1976}
Hehl F W \etal 1976 General relativity with spin and torsion: Foundations and prospects \RMP {\bf 48} 393

\bibitem{Weyssenhoff:1947}
Weyssenhoff J and Raabe A 1947 Relativistic dynamics of spin-fluids and spin-particules {\it Acta Phys. Polon.} {\bf 9} 7

\bibitem{Obukhov:1987}
Obukhov Y N and Korotky V A 1987 The Weyssenhoff fluid in Einstein-Cartan theory \CQG {\bf 4} 1633

\bibitem{Lifshitz:1946}
Lifshitz E M 1946 On the gravitational stability of the expanding universe {\it J. Phys. USSR} {\bf 10} 116

\bibitem{Bardeen:1980}
Bardeen J M 1980 Gauge-invariant cosmological perturbations \PR D {\bf 22} 1882

\bibitem{Hawking:1966}
Hawking S W 1966 Perturbations of an Expanding Universe {\it ApJ.} {\bf 145} 544

\bibitem{Ellis:1989}
Ellis G F R and Bruni M 1989 Covariant and gauge invariant approach to cosmological density fluctuations \PR D {\bf 40} 1804

\bibitem{Puetzfeld:2004}
Puetzfeld D 2004 Prospects of Non-Riemannian Cosmology
({\it Preprint} \href{http://arxiv.org/abs/astro-ph/0501231}{astro-ph/0501231})

\bibitem{Palle:1998}
Palle D 1998 On primordial cosmological density fluctuations in the Einstein-Cartan gravity and COBE data \NC  B {\bf 114} 853
({\it Preprint} \href{http://arxiv.org/abs/astro-ph/9811408}{astro-ph/9811408})

\bibitem{Ellis:1998}
Ellis G F R and van Elst H 1999 Cosmological models {\it Proceedings of the NATO Advanced Study Institute on Theoretical and Observational Cosmology, Carg\`ese, France, August 17-29, 1998 (Kluwer Academic)} {\bf 541} 1
({\it Preprint} \href{http://arxiv.org/abs/gr-qc/9812046}{gr-qc/9812046})

\bibitem{Challinor:2000}
Challinor A 2000 Microwave background anisotropies from gravitational waves: the 1 + 3 covariant approach \CQG {\bf 17} 871
({\it Preprint} \href{http://arxiv.org/abs/astro-ph/9906474}{astro-ph/9906474})

\bibitem{Maartens:1997}
Maartens R 1997 Linearisation instability of gravity waves? \PR D {\bf 55} 463
({\it Preprint} \href{http://arxiv.org/abs/astro-ph/9609198}{astro-ph/9609198})

\bibitem{vanElst:1996}
van Elst H 1996 Extensions and applications of the 1+3 decomposition methods in general relativistic cosmological modelling (PhD Thesis, Queen Mary \& Westfield College, London)

\bibitem{Hehl:1974}
Hehl F W 1974 Spin and torsion in general relativity II: geometry and field equations {\it Gen. Rel. Grav.} {\bf 5} 491

\bibitem{Palle:2007}
Palle D 2007 Comment on the paper ``Weyssenhoff fluid dynamics in a 1+3 covariant approach'' (arXiv:0706.2367v1) 
({\it Preprint} \href{http://arxiv.org/abs/0706.3144}{arXiv:0706.3144})

\end{thebibliography}
\end{document}